\documentclass{aa}

\usepackage{txfonts}
\usepackage{amsmath}
\usepackage{subfig}
\usepackage{caption}
\usepackage{hyperref}
\usepackage{multirow}
\usepackage{longtable}
\usepackage{booktabs,caption}
\usepackage[flushleft]{threeparttable}
\usepackage{tabu}
\usepackage{adjustbox}
\usepackage{array}
\usepackage{graphicx}
\usepackage{mathabx}
\newcommand{\angstrom}{\mbox{\normalfont\AA}}
\graphicspath{{./img/}}

\newcolumntype{C}[1]{>{\centering\let\newline\\\arraybackslash\hspace{0pt}}m{#1}}
\newcolumntype{L}[1]{>{\raggedright\let\newline\\\arraybackslash\hspace{0pt}}m{#1}}
\newcolumntype{P}[1]{>{\centering\arraybackslash}p{#1}}
\newcolumntype{R}[1]{>{\raggedleft\let\newline\\\arraybackslash\hspace{0pt}}m{#1}}

\begin{document} 

\title{Characterization of exoplanetary atmospheres through a model-unbiased spectral survey methodology}
 


   \author{A. Lira-Barria
          \and
          P. M. Rojo
          \and
          R. A. Mendez
          }

   \institute{Departamento de Astronomía, Universidad de Chile, Casilla 36-D, Correo Central, Santiago, Chile\\
              \email{projo@das.uchile.cl}
    }
       \date{Accepted September 12, 2021}

 
  \abstract
   {Collecting a large variety of exoplanetary atmosphere measurements is crucial to improve our understanding of exoplanets. In this context, it is likely that the field would benefit from broad species surveys, particularly using transit spectroscopy, which is the most successful technique of exoplanetary atmosphere characterization so far.}
   {Our goal is to develop a model-unbiased technique using transit spectroscopy to analyze every qualified atomic spectral line in exoplanetary transit data, and search for relative absorption, that is, a decrease in the flux of the line when the planet is transiting.}
   {We analyzed archive data from HDS at Subaru, HIRES at Keck, UVES at VLT, and HARPS at LaSilla to test our spectral survey methodology. It first filtered individual lines by relative noise levels. It also corrected for spectral offsets and telluric contamination. Our methodology performed an analysis along time and wavelength. The latter employed a bootstrap corroboration.}
   {We highlight the possible detections of Mn I and V II in HD 209459b data taken by HDS at Subaru ($5.9\sigma$ at 5916.4 $\angstrom$, $5.1\sigma$ at 6021.8 $\angstrom$). The previous detection of Ca I in the same planet is classified as inconclusive by our algorithm, but we support the previous detection of Sc II ($3.5\sigma$ at 6604.6 $\angstrom$). We also highlight the possible detection of Ca I, Sc II, and Ti II in HD 189733 data taken by UVES at VLT ($4.4\sigma$ at $6572.8 \angstrom$, $6.8\sigma$ at $6604.6 \angstrom$, and $3.5\sigma$ at $5910.1 \angstrom$), in addition to the possible detection of Al I in WASP-74b data taken by UVES at VLT ($5.6 \sigma$ at  $6696.0 \angstrom$). }
   {}


   \maketitle

\section{Introduction} 

\par Two decades after the discovery of the first exoplanet orbiting a Sun-like star \citep{1995Natur.378..355M}, and the thousands of new planet discoveries that followed, the characterization of exoplanetary atmospheres is now becoming a mature endeavor. To date, nearly one hundred exoplanetary atmospheres have been probed, almost a hundred exoplanets have at least one chemical species detected, and nearly 50 different species have been found. Exoplanetary atmosphere characterization gives us important clues to understand planetary formation and atmospheric dynamics, and it could lead to the detection of bio-signatures in the future \citep{oxygen2019}. In addition, to construct an all-encompassing planetary atmospheric theory, we need a large sample of exoplanetary atmospheres with well constrained abundances.

\par
\begin{figure*}
   \centering
   \includegraphics[width = 0.9\textwidth]{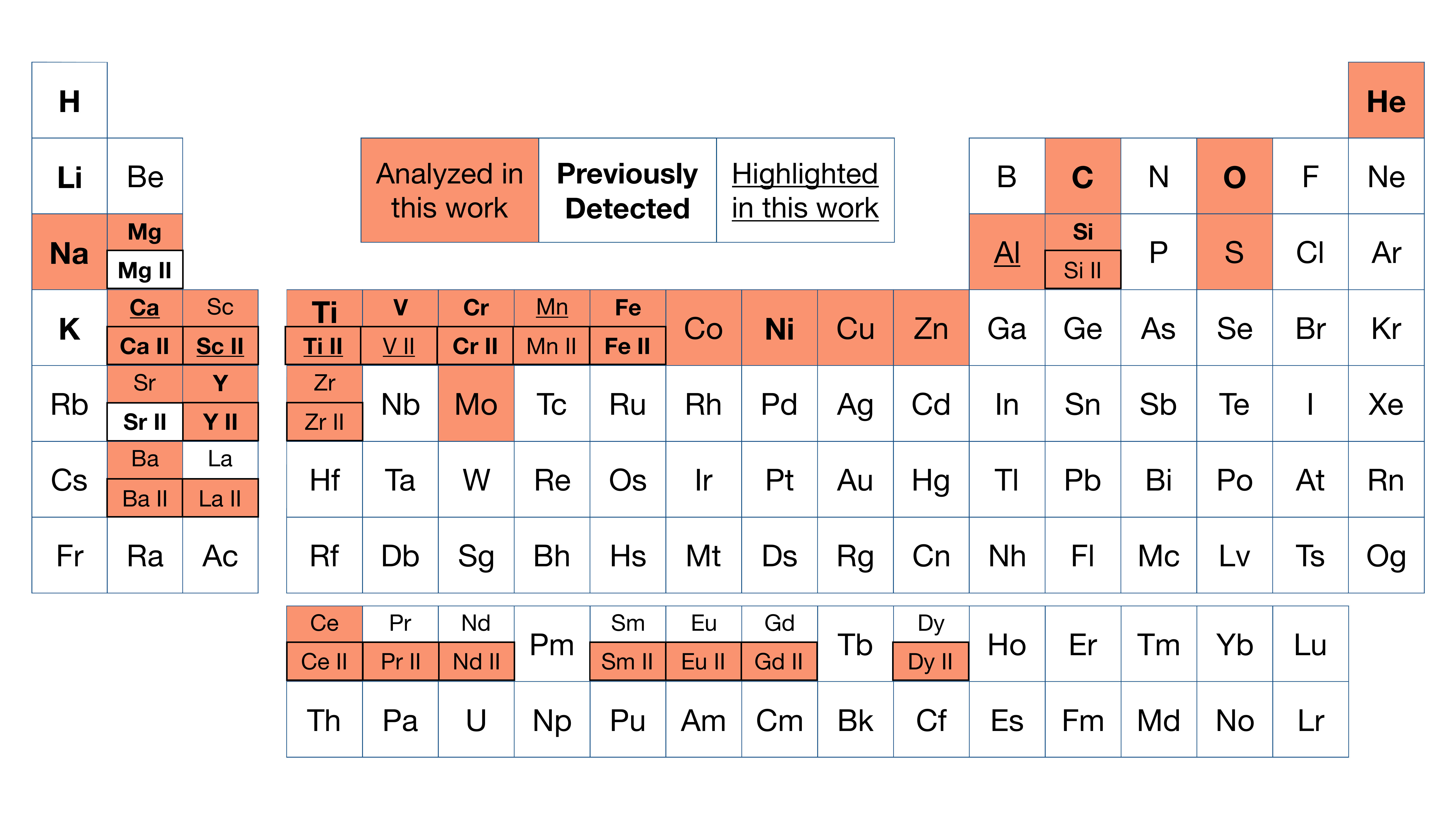}

   \caption{Periodic table of this spectral survey. In orange we indicate elements analyzed by this work, which means that at least one transition was analyzed in at least one planet. Elements reported as detected in the literature are indicated in bold. }\label{periodic_table}
\end{figure*}

 So far, transit spectroscopy has been the most successful technique to characterize exoplanetary atmospheres, both by species detected and exoplanets analyzed. This technique can be used during a primary eclipse to obtain an absorption spectrum of the planet, or during a secondary eclipse to obtain an emission spectrum. In what follows, we focus on primary eclipses. During them, transit spectroscopy takes advantage of the fact that the chemical species that are present in an exoplanetary atmosphere absorb differently as a function of  wavelength, inducing a wavelength-dependent effective planetary radius when transiting. This method has provided several molecular detections, including $H_2O$, $TiO$, and $AlO$ \citep{waterhd2094,tiowasp19,alowasp33}. Even sub-Neptunes have already been characterized using this technique \citep{waterk218}. When transit spectroscopy is performed through high resolution spectrographs, individual atomic species can be resolved  \citep[e.g., $Na, K, Mg$][]{nahd2094, nawasp49, khd1897, mgmascara}). Recently, the sample has been extended to include He I \citep{hewasp69,hewasp107, hehd2094, hehd2094} and Fe I (\citet{fekelt9}, \citet{fehd2094}, \citet{femascara2}). A complete review can be found in \citet{review2019}, while Fig.~\ref{periodic_table} shows the sample of atoms currently detected.

At temperatures above $\sim2000$~K, the emergence of the new category of ultra-hot Jupiters foresees an increase in the sample of atoms to be detected. In fact, detections of Li, Na, K, H, Ni, Ca II, Cr, Cr II, Sc II, Y II, Mg, Mg II, Fe, Fe II, Ti II, V, and possibly Ca, Co I, and Sr II have already been reported in surveys for this class of objects \citep{nawasp52, crwasp121, srkelt9, ca2+wasp121, ca2kelt9, hkelt9, ca2mascara2, mg2fe2wasp121, atomswasp121, kliwasp76, wasp121espresso}. In this context, it is likely that the field would benefit from broad spectral surveys for species. The field has already repeatedly taught us that we should expect surprises. 

An early case of a species survey was presented in \citet{adr2013}, who pioneered an unbiased blind-search algorithm. This method analyzes every atomic spectral line, not only in those positions where models predict absorption. In this work, we improve on that algorithm to uniformly analyze archival spectra from several instruments, including the High Dispersion Spectrograph (HDS) on the Subaru telescope, the High Resolution Echelle Spectrometer (HIRES) on the Keck telescope, the Ultraviolet and Visual Echelle Spectrograph (UVES) on the Very Large Telescope (VLT) and the High Accuracy Radial Velocity Planet Searcher (HARPS) on the ESO 3.6 m in La Silla. Our objective was to expand the sample of species detected, using data of transits mostly taken to perform Rossiter-McLaughlin measurements \citep{rossiter1924, mclaughlin1924}, as they need a similar observing setup as transit spectroscopy observations. Our methodology contributes to the field, making transit spectroscopy surveys easier since it will be freely available to be used by anyone. We already have some tentative detections, and thanks to the versatility of the code, many other planets can be analyzed with it in a straightforward way.

This work is organized as follows. We introduce our methodology in Sect.~\ref{algorithm}, giving details for all the steps of our algorithm. We then describe the observations in Sect.~\ref{observations} and present our main results in Sect.~\ref{results}. We conclude in Sect.~\ref{conclusions} with an outline of future work and elaborate on how this technique could help the field.

\section{Methodology\label{algorithm}}

\begin{figure*}[ht]
   \centering

   \includegraphics[width = 16cm]{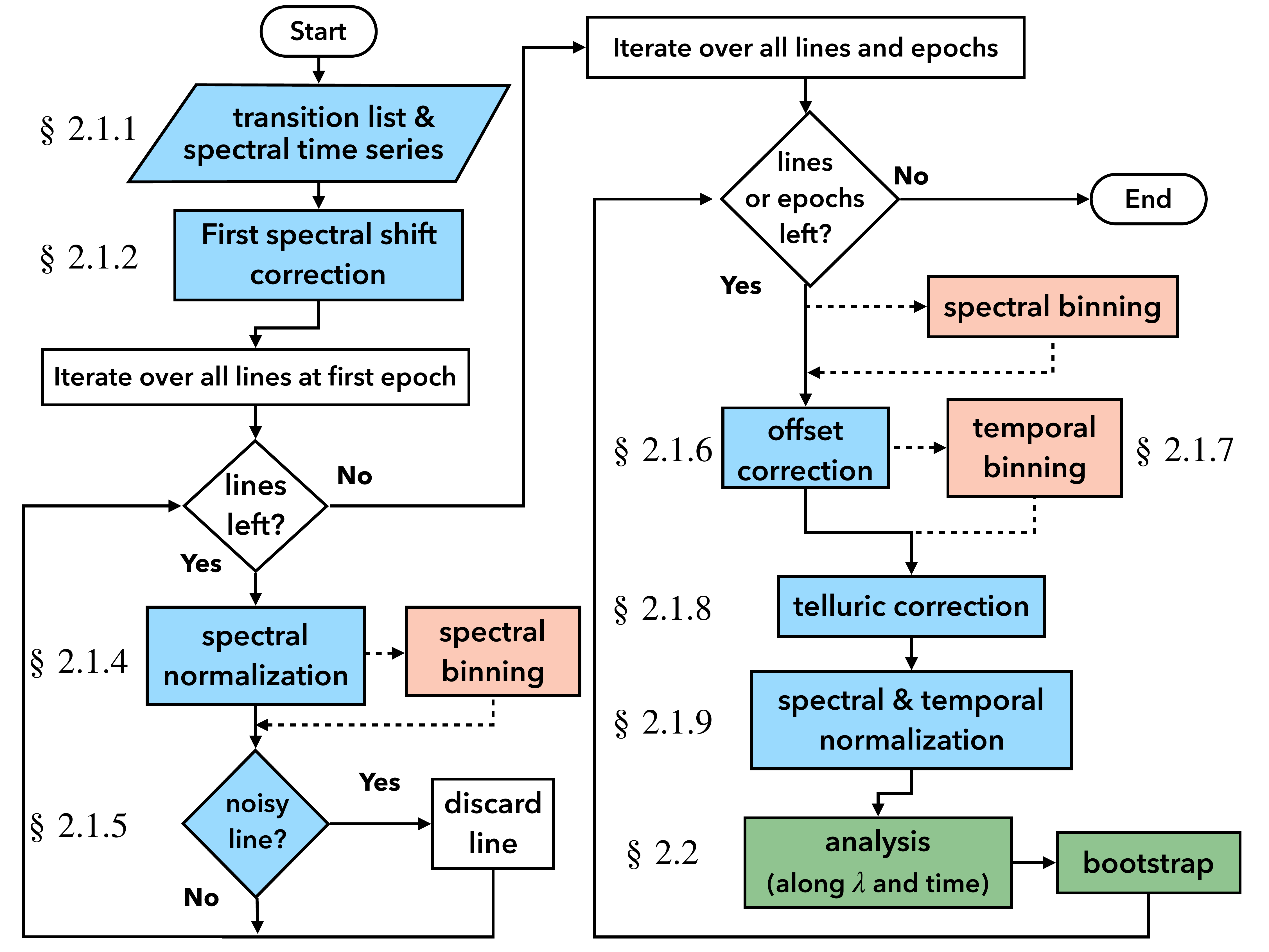}
   \caption{Flowchart of our methodology.}
              \label{fig: method}
\end{figure*}

Our method has been entirely coded on Python 3.7, and it is currently available on github\footnote{https://github.com/alirabarria/TotalBlindSearch}.  Multiprocessing has been implemented and most of the numerical parameters mentioned below can be customized. Our algorithm can be used to perform a blind (model-unbiased) search for absorption features over the entire wavelength range of transit spectral time series, but it can also be used to perform a directed search for one particular species. A typical transit observed with an optical echelle spectrograph can be analyzed in a few days using twenty cores. The flowchart of the algorithm is shown in Fig.~\ref{fig: method}.

\par Our approach is mainly based on the methods described by \citet{adr2013} (ADR2013 hereafter). The algorithm is designed to analyze every transition approved by our qualifying tests (QT, see Sect.~ \ref{data_preparation}.\ref{tests}) to search for relative absorption. This is achieved by comparing zones where the flux should remain constant with those where a relative decrease is expected, both along time and wavelength. The latter is studied with a passband centered at the transition, which is compared with the nearby continuum. Along time, the algorithm compares in- and out-transit data. Finally, to calculate the confidence of any possible absorption, the code takes several random samples (using a bootstrap approach) of the time series. It compares cases when only a few in-transit data are ignored (to avoid possible outliers), with cases without the relative absorption. The latter is achieved by only taking out(in)-transit data, to calculate the relative decrease in flux (see Fig.~\ref{fig:scenarios}). If a particular species is present in the exoplanetary atmosphere, a consistent detection should exhibit absorption in several of its qualified transitions. The algorithm steps are described in the following subsection.

\subsection{Data preparation and corrections}
\label{data_preparation}

\begin{enumerate}

\item To localize the known transitions, we used the entire solar Spectroweb transition list \citep{spectroweb}. We skipped all molecular transitions and, differently from ADR\citeyear{adr2013}, no line depth filter was applied. We also added the He I transitions from the Vienna Atomic Line Database (VALD) \citep{vald, vald2, WSG}. \footnote{The original source for some of these lines is http://kurucz.harvard.edu/linelists.html} 

\item We made a first spectral shift correction by considering the published systemic radial velocity of the star. We iteratively refined this shift by considering a selection of strong features ($\log{gf} > -1$ , solar line depth $> 20 \%$) discarding outliers, until the scatter on the residuals was $\leq 0.2 \angstrom$. 

\item For all the following steps, we used a reduced local wavelength range of $ 60 \times FWHM$ centered at each transition provided by the database. 

\item To normalize the spectra, our first procedure was to apply a median filter using a kernel of 5 $\times$ the empirical full width at half maximum (FWHM). Continuum pixels (CP) \label{cp} were selected within 3.5 Poisson-statistic sigmas from the median filter. Then, we fit a third order polynomial to the CP, which was used to divide the spectra by it. At this step, Poisson statistics were calculated using the telescope gain and the counts; only the UVES’ pipeline provided a more detailed and precise noise characterization, which included the sky and dark subtraction. 

\item \label{tests} To discard noisy or blended lines, we tested the transitions by fitting a Gaussian profile to the first frame of each observed transition. Accepted lines had to pass the QT described in the following steps:

\begin{itemize}
     \item  Amplitude: The amplitude of the fit must be closer than 6\% to the normalized minimum flux.
     \item  Center: The wavelength at the minimum flux must be closer than 0.06 $\angstrom$ from the center of the fit.
     \item Residuals: At least 70$\%$ of CP must be closer than 2$\times$ the expected sigma noise\footnote{Calculated from the empirical S/N (see Table \ref{tab:data}).} to the median of the CP.
     
 \end{itemize} 

\item To correct local systematic offsets between the same transition at different frames. \footnote{The following four steps are modifications with respect to ADR\citeyear{adr2013}.} We used cross-correlation to locally align the frames. For each local wavelength range on each frame, the algorithm tested 2500 intervals in a range of $\pm 1 \angstrom$ and always compared the local range to the interpolated flux of the first frame. 

\item To increase the signal-to-noise ratio (S/N), our pipeline can average data along wavelength or time. Wavelength binning is carried out before normalization, while time binning is carried out after performing the misalignment correction. For this work, we did not use this feature for all planets (see Table \ref{tab:data}). 

\item To calculate a telluric spectrum, we fit the intensity to the airmass $I = I_0 \exp{(Nk_\lambda s)}$, where $I$ is the intensity at a certain wavelength $\lambda$, $I_0$ is the source intensity, $Nk_\lambda$ is the optical depth at zenith, and $s$ is the airmass. We used an exponential model in order to maintain error consistency\footnote{ADR\citeyear{adr2013} used a linear model to the logarithm of the flux, which was not statistically correct.}. 

\item After performing the misalignment and telluric correction, we renormalized each line at each frame. Then, following \citet{hoijmakers2014}, we divided each pixel by its median in time. Since the planet signal shifts due to the planetary orbital motion, we kept the planet signal but removed the stellar spectrum. 

\item We finally shifted all frames to the planetary rest frame. We calculated the shift from the ephemeris, semi-major axis, and period of each planet, assuming it follows a circular orbit. 

\end{enumerate}

\subsection{Analysis}

The algorithm's main objective is to search for relative absorption in flux on different domains. A robust detection shows a match along time and along wavelength. To assess the former, following \citet{snellen2008}, we created a passband centered at the transition and two continuum passbands at the sides (we tried lengths of 0.75 and 1 $\angstrom $ for every object analyzed). We averaged the bands and compared the flux at the feature’s center with the surrounding passbands to calculate the relative absorption at each transition. If absorption is present, we should see a transit-shaped curve along the time axis. We used this criteria as a qualitative test of the result. 

\begin{figure*}[ht]
   \centering
   \includegraphics[width = 0.9\textwidth]{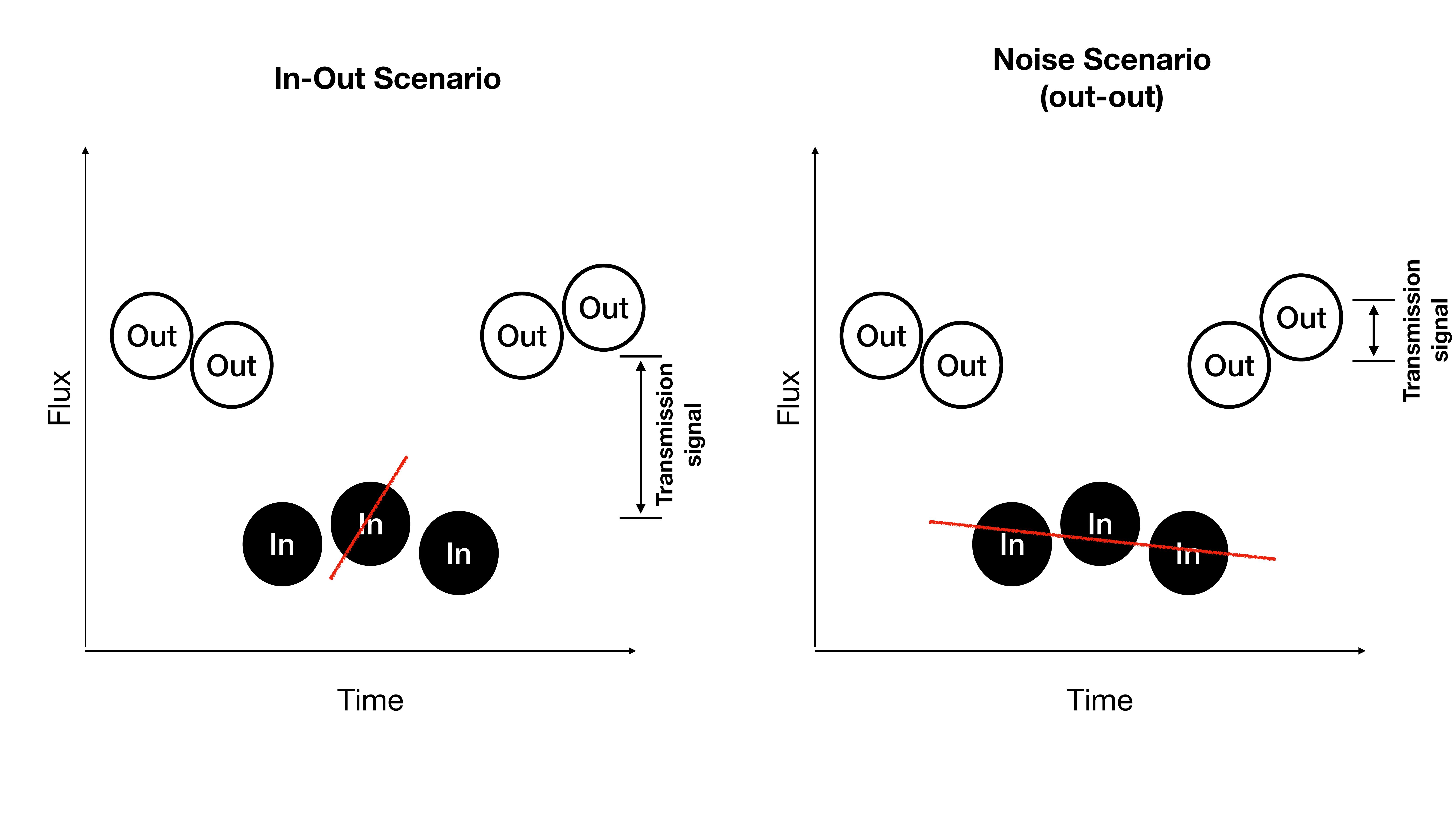}
  
\caption{Graphical representation of our analysis along wavelength, using bootstrap scenarios adapted from \citet{redfield2008}. \textit{The In-Out scenario:} In order to avoid the situation where a few frames might be dominating the absorption, we removed an increasing number of in-transit frames at random to calculate the relative absorption (starting with two and ending at half of the sample). \textit{Noise scenarios:} For the Out-Out scenario, we ignored all the in-transit frames and labeled some of the remaining frames as in-transit to calculate the absorption. We did the same for the In-In scenario as the Out-Out one, but we ignored the out-transit frames. The distribution of the In-Out scenario is expected to be centered at the actual relative absorption, while noise scenarios should be centered at zero. We used the 1$\sigma$ value of the In-Out histogram as our uncertainty. Additionally, 3000 iterations of these scenarios were made for each transition. However, when the frames were scarce, we made the maximum number of iterations allowed without repeating possibilities.}\label{fig:scenarios}
\end{figure*}

On the other hand, to assess the detection along wavelength, we obtained the transmission spectrum by averaging the in- and out-of-transit frames and then calculating $(F_{In} - F_{Out})/F_{Out}$ at each pixel. If the transition shows absorption, we should see a Voigt-like profile. Following \citet{redfield2008}, and using the $0.75$ and $1 \angstrom$ passbands from the previous analysis, we bootstrapped three types of scenarios to estimate uncertainties. For each scenario, the relative absorption was calculated by subsampling data to search for false positives and to test the robustness of a detection. The In-Out scenario tests whether the result is robust against a few frames biasing the absorption, while the Out-Out and In-In scenarios test against the possibility of random noise generating the signal (see Fig.~\ref{fig:scenarios} for details of the bootstrap algorithm). We used this criteria as a quantitative test of the result.

\section{Observations\label{observations}}
 
We only considered archival data, including all planets with Rossiter-McLaughlin measurements. Additionally, we included a few other planets with archival data. We used a set of prominent transitions to test every object, and we discarded the data if none of these transitions would pass the QT (see Sect.~ \ref{data_preparation}.\ref{tests}). Out of our 19 initial targets, only six targets remained as shown in Table~\ref{tab:data}. In addition, we focused only on the systems with the highest expected S/N computed from the equilibrium temperature, flux of the star, and diameter of the telescope. The ephemeris of each target is shown in Table~\ref{tab:eph}.

Our selected datasets were mostly observed by telescopes with diameters greater than 8m, with the ESO 3.6m telescope being the only exception. We only used data from high-resolution spectrographs in order to resolve atomic transitions. They all lie in the 50000-70000 range, except for HARPS that has a resolving power of 110000. 

For HARPS, after testing a few objects (WASP-167b, GJ 436 b), we realized that for objects fainter than $\sim 10.5$~mag, the signal was too weak to be analyzed by our method, as no transitions passed the QT. However, for HIRES, we could analyze down to magnitude $11.41$, corresponding to target TrES-2b. 

HIRES data were reduced automatically using the MAuna Kea Echelle Extraction (MAKEE) pipeline \citep{makee}. HARPS and UVES science data were obtained from the ESO archive, and the HDS data were already reduced by \citet{adr2013}. This previous work applied a nonlinearity correction that \citet{snellen2008} suggested for HDS. We used the data already corrected by them and we did not analyze other objects observed by this spectrograph due to the lack of this correction. \citet{hd2094uves} suggested a nonlinearity correction for an effect $\sim$3 times smaller in UVES data than in HDS. \citet{hd1897uves} did not apply this correction, and neither did we. Additionally, we discarded the UVES data set taken by \citet{hd2094uves} because they lacked observations of pre-transit time.

 Based on our transition lists, we concentrated on the bands that fit in the wavelength range from $\sim$5000 to $\sim$7000$\angstrom$. This corresponded to the red CCD on the HDS dataset, the red band on the UVES dataset\footnote{Despite the fact that it covers until $9400 \angstrom$, our analysis only covers until $\sim 7000 \angstrom$.}, and the green CCD on the HIRES dataset. In the latter, we also noted that its noise level was low compared to the blue CCD of the same spectrograph. On the other hand, the HARPS dataset was analyzed entirely.

\begin{figure*}[ht]
   \centering
    \includegraphics[width=0.9\textwidth]{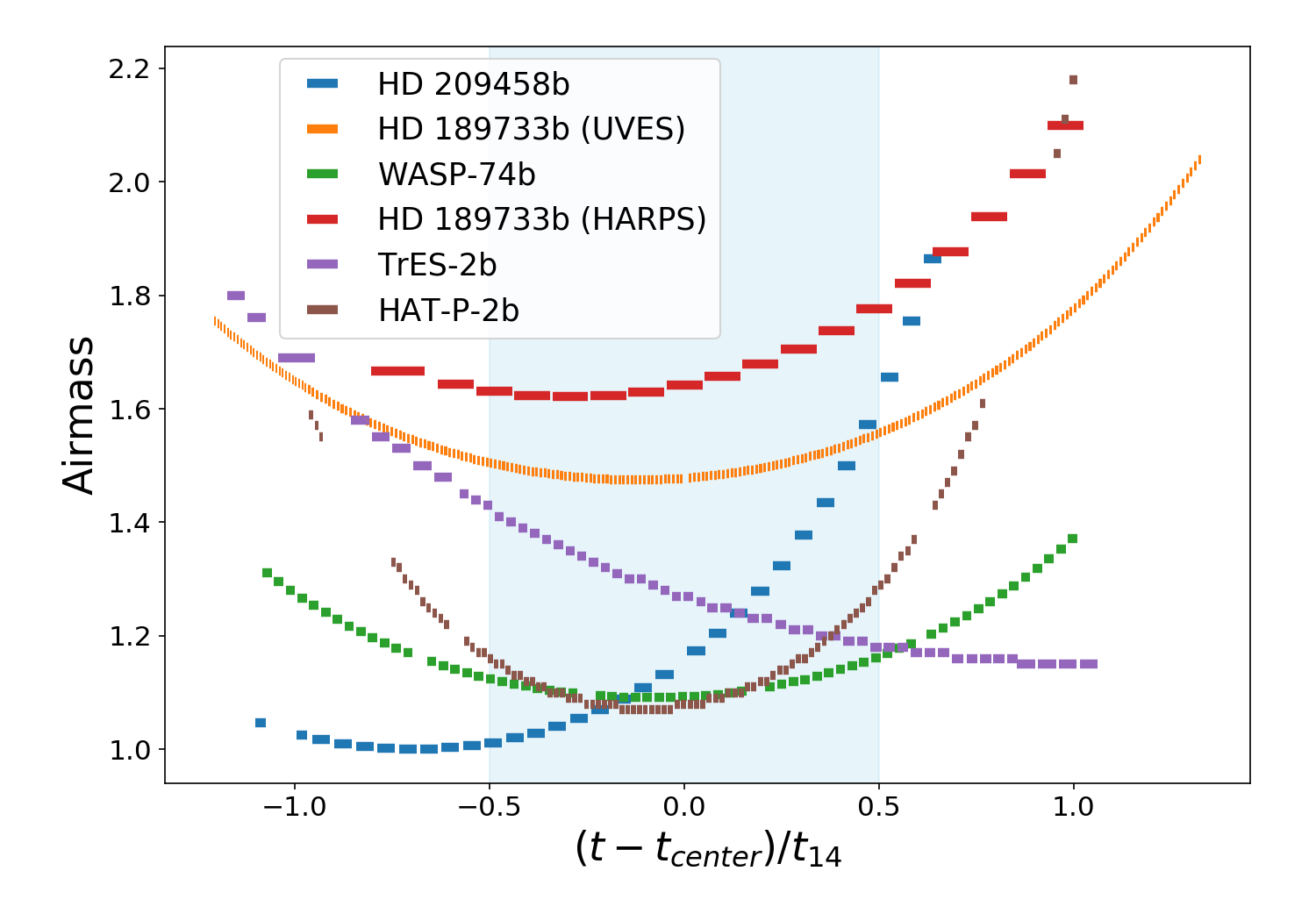}
    
   \caption{Transit coverage of all planets analyzed in this paper. Each line represents a frame, and the length of the line was calculated using the exposure time adopted at that frame. It is important to note the dispersion of exposure times. The transit duration is shown as a light blue band.}
              \label{fig:coverage}%
\end{figure*}

From Table~\ref{tab:data} we see that HD~209458b at HDS has a S/N per frame that is considerably larger than every other dataset. In order to improve the S/N of the other objects analyzed, we binned in wavelength and/or time if possible. Wavelength binning was only performed if the number of pixels per $\angstrom$ was bigger than HDS’, while time binning was only applied if we could maintain the number of in-transit frames similar in all analyzed objects ($\sim$20), as we need a similar number of combinations to perform the bootstrap analysis. The latter binning was mostly uniform, discarding frames right before or after the transit when necessary. TrES-2b was a special case because we corrected the changing exposure time, resulting in a nonuniform binning of frames. A comparison of transit coverage and exposure time for each target is shown in Fig.~\ref{fig:coverage}.

\begin{table*}[!ht]
\caption{Archival spectral data analyzed in this paper. The S/N was calculated empirically from the residuals on the continuum of transitions approved by our QT. The S/N per frame was calculated in each frame and scaled by the number of pixels per angstrom. On the other hand, the S/N per transit was calculated on the first frame and scaled by the number of pixels in one FWHM and the number of in-transit frames ($\#$in frames at the table). In the binning column, 2s means that we binned two pixels in wavelength, and 4t means that we binned four frames in time. TrES-2b was a special binning case, as we performed a nonuniform time binning. The references for each dataset are listed below the table.}
\begin{threeparttable}
\begin{tabular}{llllllll}
 \hline \hline
Object      & Date       & Instrument    & S/N (per frame) & S/N (transit) & $\lambda$ range (nm) & $\#$In frames & Binning \\
 \hline
HD 209458b \textsuperscript{1}& 10-24-2002 & HDS at Subaru & 1300 & 2300 & 550-680 & 17 & No \\
HD 189733b \textsuperscript{2}& 07-01-2012 & UVES at VLT & 110 & 2500 & 560-940 & 92 & 2s, 4t \\
HD 189733b \textsuperscript{3}& 09-07-2006 & HARPS at LaSilla & 100 & 770 & 380-690 & 19 & 2s \\
TrES-2b \textsuperscript{4}& 04-26-2007 & HIRES at Keck & 200 & 660 & 490-640 & 30 & $\sim$ 4t \\
WASP-74b \textsuperscript{5}& 07-28-2017 & UVES at VLT & 80 & 1400 & 560-940 & 14 & 2s \\
HAT-P-2b \textsuperscript{6}& 06-06-2007 & HIRES at Keck & 300 & 2500 & 490-640 & 62 & 3t \\

 \hline
\end{tabular}
\begin{tablenotes}
      \small
      \item \textit{References:} 1: ADR\citeyear{adr2013}, 2: \citet{hd1897uves}, 3: 072.C-0488(E), 4: \citet{datatres2},
      5: \citet{wasp74eph}, 6: \citet{datahatp2}
    \end{tablenotes}
\end{threeparttable}
\label{tab:data}
\end{table*}

\begin{table*}
\caption{Ephemeris of each planet analyzed in this work. The references listed correspond to ephemeris and transit duration. If the former is lacking, then the reference of the ephemeris and the transit duration is the same. We used the same values of ephemeris for both datasets of HD 189733 b.}
\begin{tabular}{lllll}
 \hline \hline
Object     & $T_C$ & P & $T_{14}$ & References \\
 \hline
HD 209458b & 2452826.6285 $\pm$ 0.000087 & 3.52474 $\pm$ 0.00000038 & 2.978 $\pm$ 0.051   & \vtop{\hbox{\strut \citet{hd2094eph}}\hbox{\strut \citet{hd2094dur}}} \\
HD 189733b & 2454279.436714 $\pm$ 0.000015 & 2.21857567 $\pm$ 0.00000015 & 0.07527 $\pm$ 0.00037 & \vtop{\hbox{\strut \citet{hd1897eph}}\hbox{\strut \citet{hd1897dur}}} \\
TrES-2b    & 2453957.63479 $\pm$ 0.00038  & 2.470621 $\pm$ 0.000017   & 0.07408 $\pm$ 0.0008  & \citet{tres2eph}                     \\
WASP-74b   & 2456506.8918 $\pm$ 0.0002   & 2.13775 $\pm$ 0.000001   & 0.0955  $\pm$ 0.0008  & \citet{wasp74eph}                    \\
HAT-P-2b   & 2454387.49375  $\pm$ 0.00074  & 5.6334729 $\pm$ 0.0000061  & 0.1787 $\pm$ 0.0013  &    \citet{hatp2beph} \\             \hline
\end{tabular}
 \label{tab:eph}
\end{table*}


\section{Results\label{results}}

\begin{figure*}[t]
   \centering
    \subfloat{
    \includegraphics[width = 0.5\textwidth]{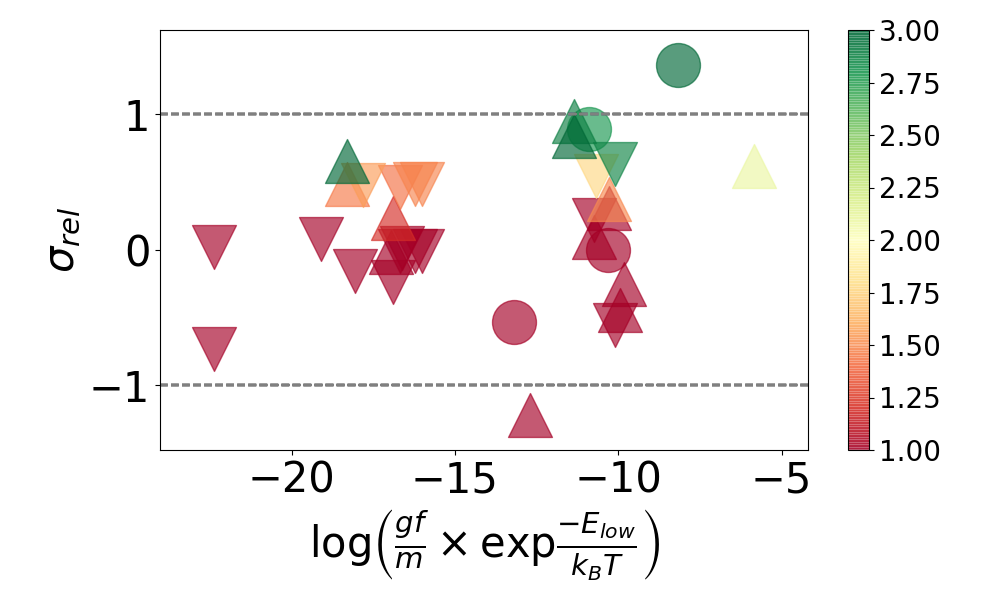}
    
    \includegraphics[width = 0.5\textwidth]{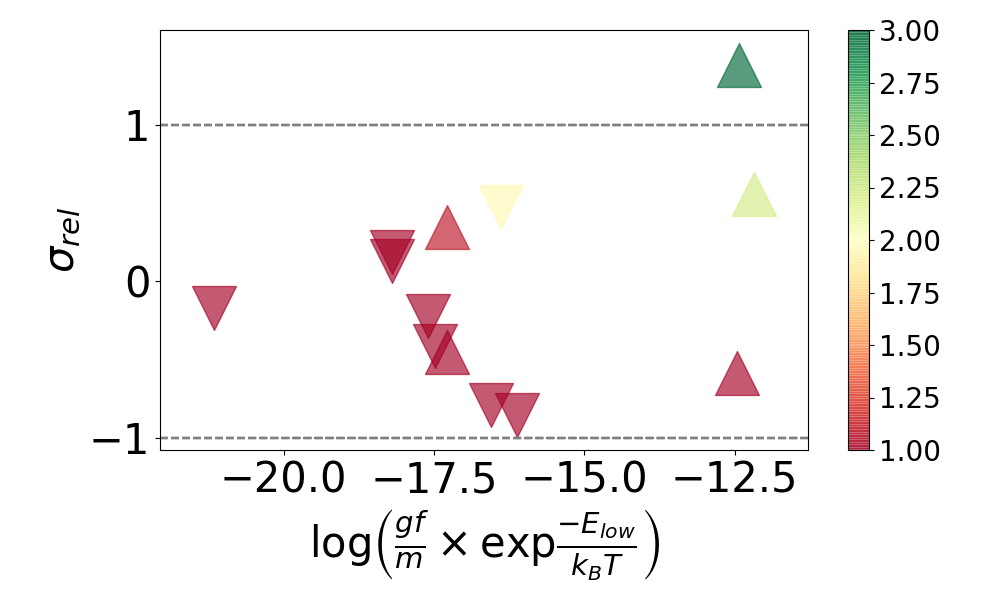}
    }
    \vfill
    \subfloat{
    \includegraphics[width = 0.5\textwidth]{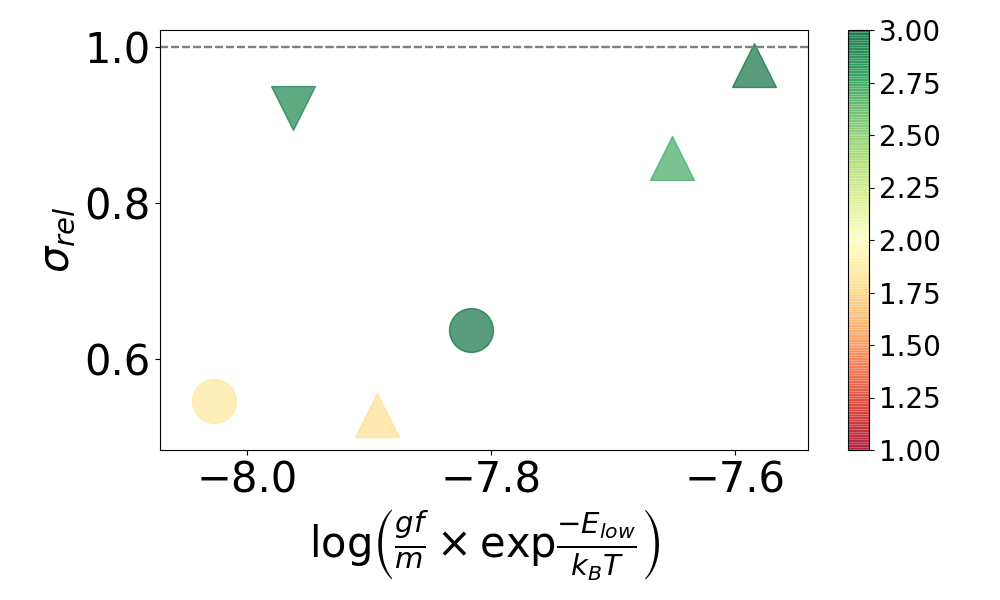}
    
    \includegraphics[width = 0.5\textwidth]{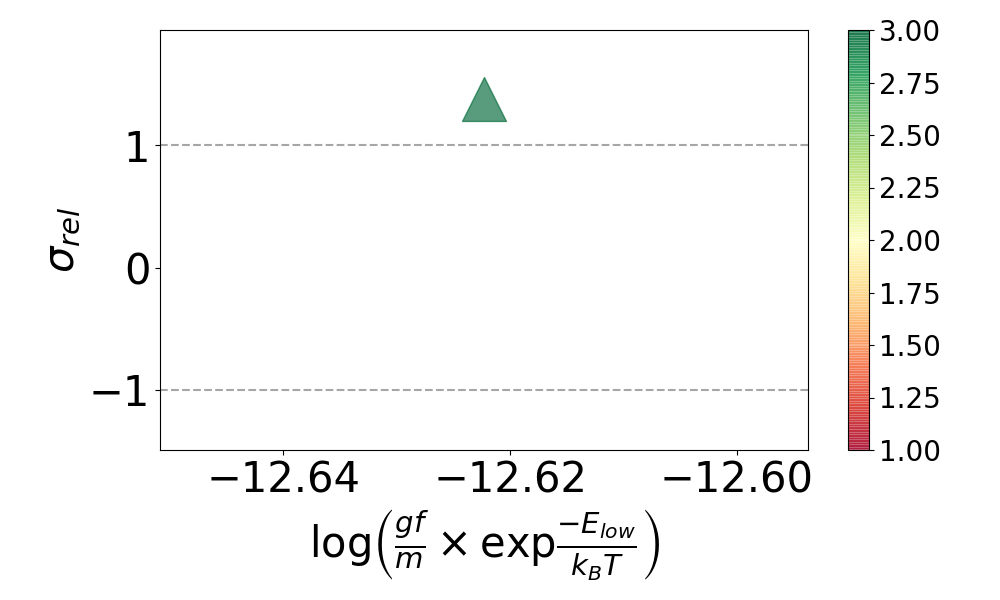}
    }
   
   \caption{Element absorption in the 0.75 $\angstrom$ passband of all the transitions approved by QT, for selected elements in HD 209458b. (\textit{Upper left:} Ca I, \textit{Upper right:} Mn I, \textit{Bottom left:} Sc II, \textit{Bottom right:} V II). The ordinate shows $(- C_{IO} )/ \Sigma_{noise}$ where $\Sigma_{noise} = max(\Sigma_{II}, \Sigma_{OO})$. The color shows the detection significance ($-C_{IO} / \Sigma_{IO}$), where green is $ \geq 3\sigma$ and red is $\leq 1 \sigma$. Additionally, to deal with contamination from transitions of other species closer than 1$\times$FWHM, we show the ones without contamination as circles ($\bullet$), with the strongest oscillator strength at that element as triangles pointing up ($\blacktriangleup$). Both types are considered strong transitions (ST). Otherwise, we show them as triangles pointing down ($\blacktriangledown$), and considered them as weak transitions.}
              \label{fig:atom_hd20}%
\end{figure*}

For each planet, we used diagnostic plots such as the one shown in Fig.~\ref{fig:atom_hd20} to examine the overall behavior of all the transitions in an atom. Then, we looked at individual transitions in detail (e.g., Fig.~\ref{fig:valid}). The latter kind of plot contains all the information of the relative absorption, depicted along wavelength and time, as well as the final bootstrap distributions (see Fig.~ \ref{fig:scenarios}).

To decide if an atomic species\footnote{We use the terms element, atom, or species interchangeably.} was detected, we produced a uniform analysis of every transition approved by the QT for each atom. From the histogram obtained for the In-Out distribution, we obtained the detection significance of the transition $\sigma_{In-Out}$ as the distance from the In-Out histogram center $(C_{IO})$ to zero, divided by the width of the In-Out histogram $(\Sigma_{IO})$ \footnote{$\Sigma$ is used for values which are in transmission spectrum units, while $\sigma$ is used for values relative to the certainty.}. Additionally, to reduce the possibility of detecting species by chance, we required that $C_{IO}$ did not remain inside a noise distribution. To do that, we considered the magnitude of the systematic noise at the line by defining $\sigma_{rel}$ as the distance between $C_{IO}$ and zero divided by the biggest noise distribution width:

$$    
\sigma_{rel} = \frac{-C_{IO}}{max(\Sigma_{II}, \Sigma_{OO})} 
$$
where $\Sigma_{II}$ and $\Sigma_{OO}$ are the width of the In-In and Out-Out histograms, respectively. We considered a transition as detected only if $\sigma_{rel} > 1$, and $\sigma_{In-Out} > 3$.

ADR\citeyear{adr2013} used the oscillator strength $(gf)$  as a reference for how strong each spectral line is. Following the absorption coefficient dependencies, we improved this proxy by adding the energy of the lower level of the transition $(E_{low})$ and the atomic mass $(m)$, and we represented the temperature of the atmospheric layer with the equilibrium temperature $(T)$ of each planet. Our new proxy for the strength $(s)$ of the line is:

$$
s \, \propto \, \frac{gf}{m} \exp {\left( -\frac{E_{low}}{k_BT} \right) }
$$
where $k_B$ is the Boltzmann constant.

In addition, we searched for contamination from other transitions closer than 1$\times$FWHM. We considered a spectral line as a strong transition (ST) if it does not share the range with any other spectral line or if it is the strongest one in that range (which we show as $\bullet$ and $\blacktriangleup$ in Fig.~\ref{fig:atom_hd20}, respectively). Otherwise we considered it is a weak transition (which we show as $\blacktriangledown$ in Fig.~\ref{fig:atom_hd20}). Additionally, in our analysis we included the noise of the continuum used to normalize the transition, defining $Sect.~igma_c$ as the standard deviation of the CP (see Sect.~ \ref{data_preparation}.\ref{cp}) at each spectral line.

Table~\ref{tab:species} summarizes our findings for all species in all of our planets analyzed, including a literature comparison. In the following sections, we highlight the interesting species found in each planet analyzed. We listed the relevant parameters of the first ten ST of selected atoms of HD 209458b in Table~\ref{tab:elhd20}.

\begin{table*}[t]
    \centering
    \begin{tabular}{ccccc}
    \hline\hline
Object & Highlighted & Rejected & Literature \\
\hline
HD 209458b & Mn I, V II & S I & H I, He I, C I, O I, Mg I, Fe II,  Na I ?, Ca I, Sc II \\
HD 189733b & Ca I,  Sc II, Ti II &  & H I, He I,  Na I, O I, \\
WASP-74b & Al I &  & No previous detections \\
    \hline
    \end{tabular}

    \caption{Summary of highlighted elements in all objects analyzed.}
    \label{tab:species}
\end{table*}

\subsection{HD 209458b (HDS at Subaru)}

Mn I shows an interesting absorption at $6016.7 \angstrom$, with a $5.9 \sigma$ of detection significance ($\Sigma_{IO}$) on the $0.75 \angstrom$ passband. Despite the fact that this could be a promising detection, since this transition is the second of importance for this element according to our proxy (and more than three orders of magnitude stronger at Mn I when compared to any other element transitions in its closer wavelength range), no other transition shows absorption in this element, including the strongest transition. Another interesting element is V II, in which only $5916.4 \angstrom$ approved the QT. This transition shows $5.1 \sigma$ of detection at the $0.75 \angstrom$ passband. However, since this is based on only one transition, it is insufficient to conclude that it is a robust detection. Therefore, we think that V II along with Mn I should be looked at carefully in future searches of atoms in this planet.

\begin{figure*}[t]
   \centering
   \subfloat{
   \includegraphics[width = 0.7\textwidth]{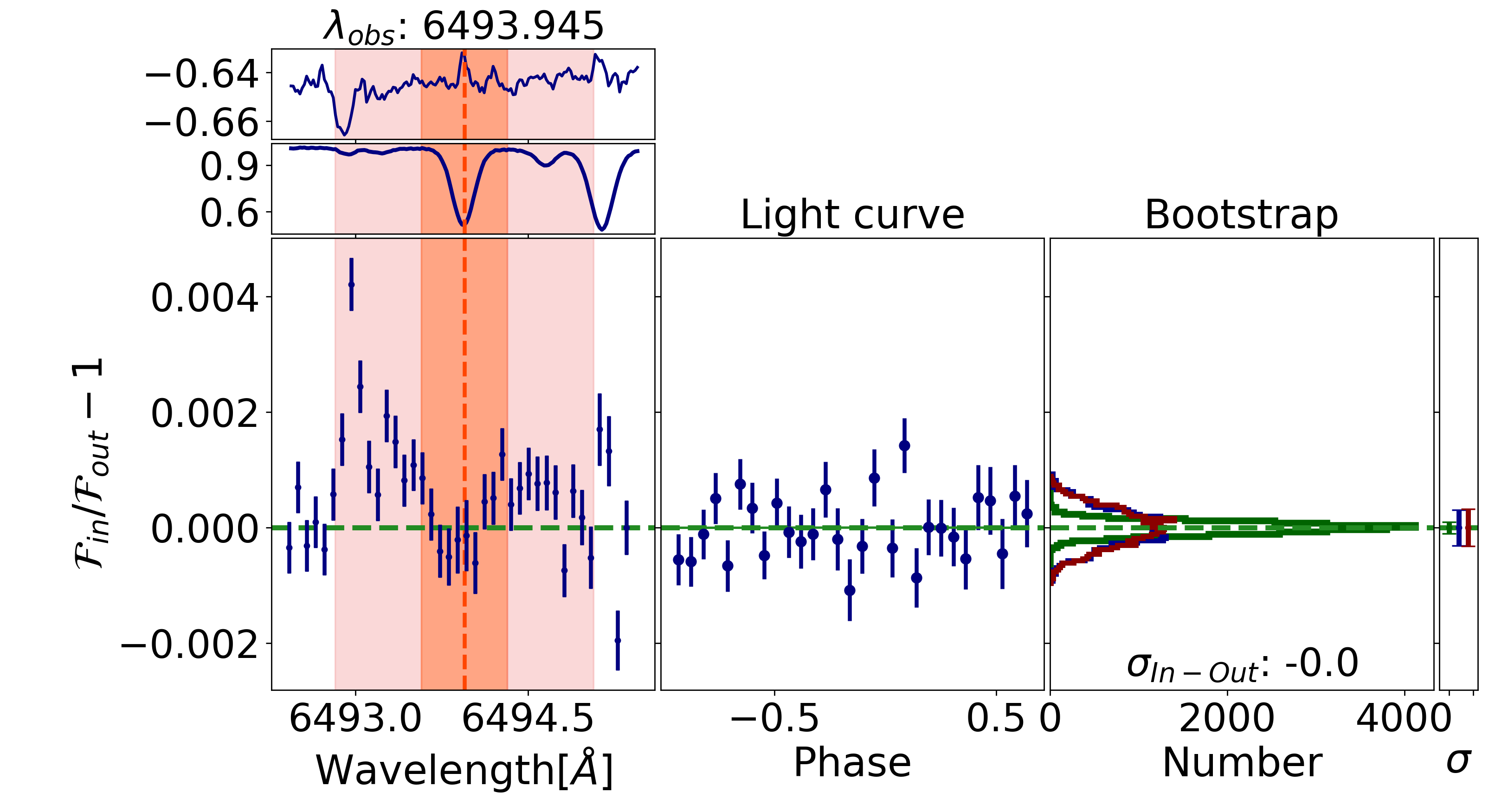}
   }
    \vfill
    \subfloat{
      \includegraphics[width = 0.7\textwidth]{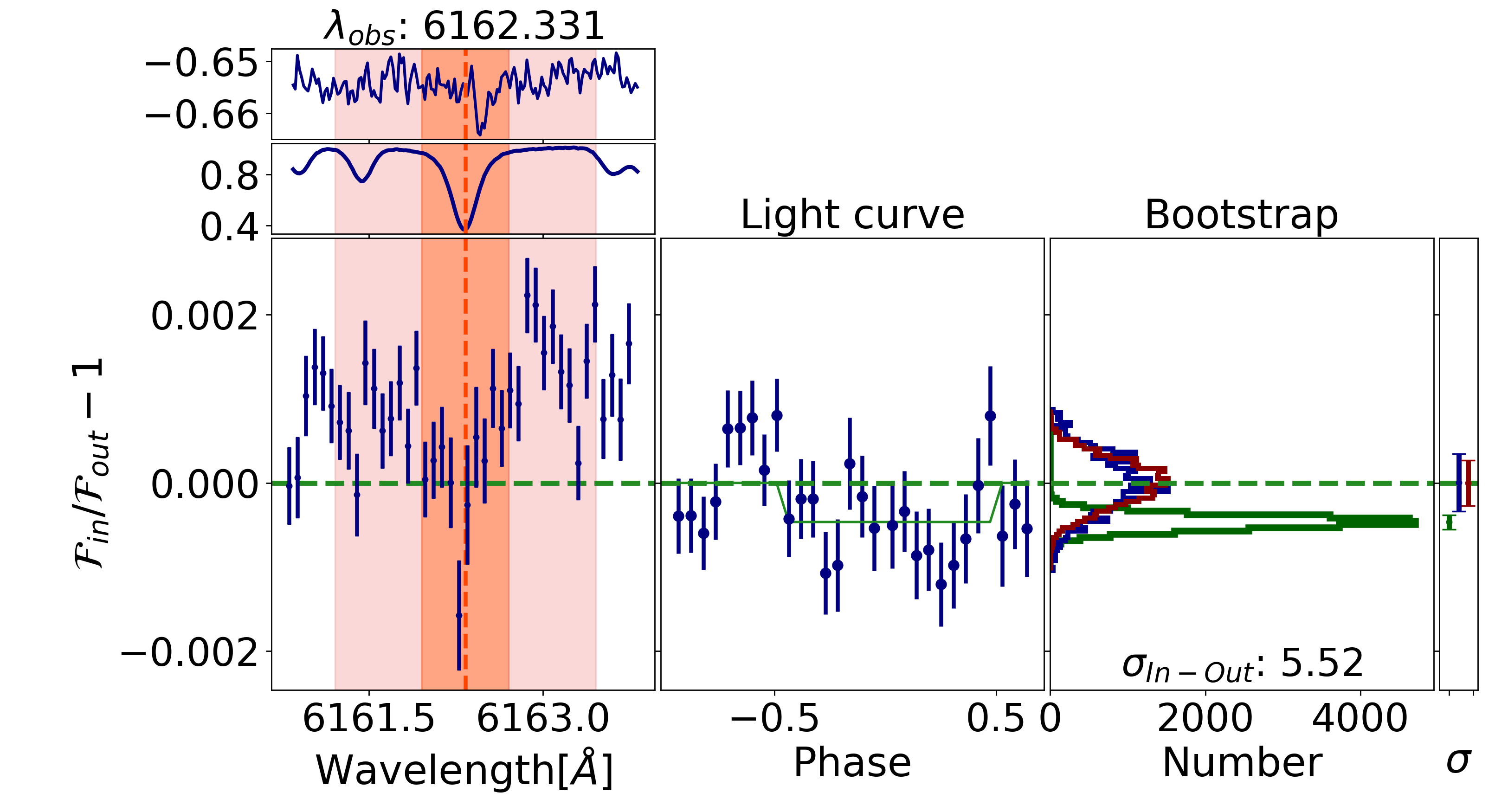}
    }
    \vfill
    \subfloat{
      \includegraphics[width = 0.7\textwidth]{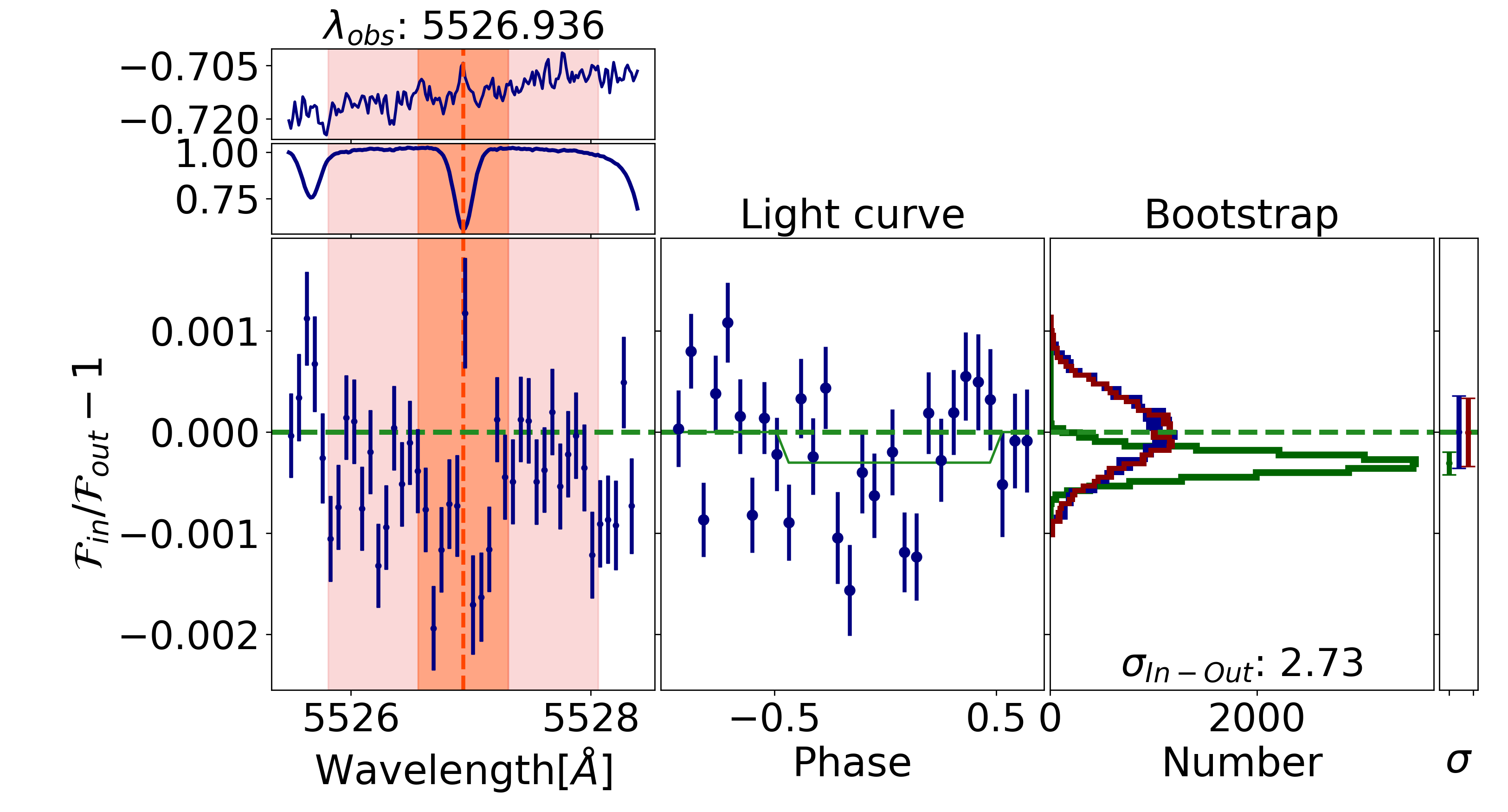}
   
   }
   
   \caption{Summary plots for previously detected Calcium I ($6493.8, 6162.3\angstrom$) and Scandium II ($5526.8 \angstrom$) revisited in this work. A summary plot is associated with a transition, and contains the principal outputs of the algorithm steps, in three panels. \\ \textit{Left panel:} From top to bottom, we see the telluric spectra, the normalized flux of the line, and the transit seen along wavelength, as the flux is integrated along time. \textit{Central panel:} \citet{snellen2008} transmission light curve. \textit{Right wide panel:} Bootstrap method plot, where we can see the distribution of the scenarios explained in Fig.  \ref{fig:scenarios}. Green corresponds to the In-Out scenario, blue to the Out-Out scenario, and red to the In-In scenario. We can also see the widths of the histograms on the right. \textit{Right narrow panel:} The graphical representation of the detection significance (green bar) and the null hypothesis (blue and red bars). The green bar corresponds to the In-Out histogram width, and the blue and red correspond to the Out-Out and In-In histograms, respectively.}
              \label{fig:valid}%
\end{figure*}

Regarding the previous Ca I detection reported by ADR\citeyear{adr2013}, we found evidence of absorption in some of their detected transitions. However, the values of $\sigma_{In-Out}$ were lower than in the previous work, and we did not find evidence of absorption at the $6943.8 \angstrom$ line (see Fig.~\ref{fig:valid}). Additionally, according to our proxy values, there are two ST that stand out. The strongest transition ($6572.8 \angstrom$) did not show detection, but it exhibits too much structure in its continuum passbands and in its telluric spectrum. The second ST ($6162.2 \angstrom$) shows some structure in its continuum passbands as well, but it showed a robust absorption, and it is more than one order of magnitude stronger than any other transition of Ca I. In summary, we find that the previous detection of this element (Fig.~\ref{fig:valid}) is inconclusive.

\begin{figure*}[t]
    \centering
   \subfloat{
   \includegraphics[width = 0.7\textwidth]{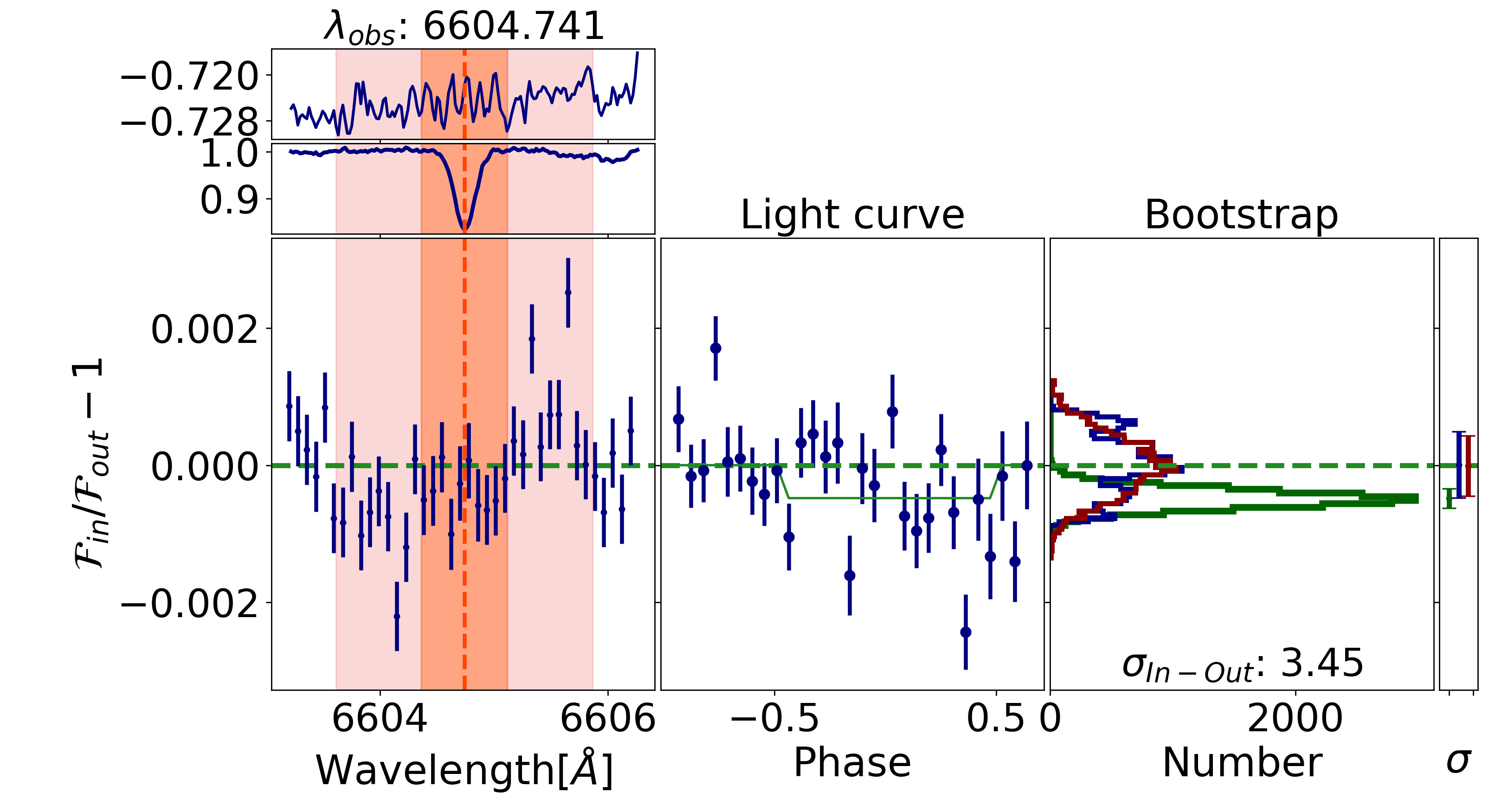}
   }
    \vfill
    \subfloat{
      \includegraphics[width = 0.7\textwidth]{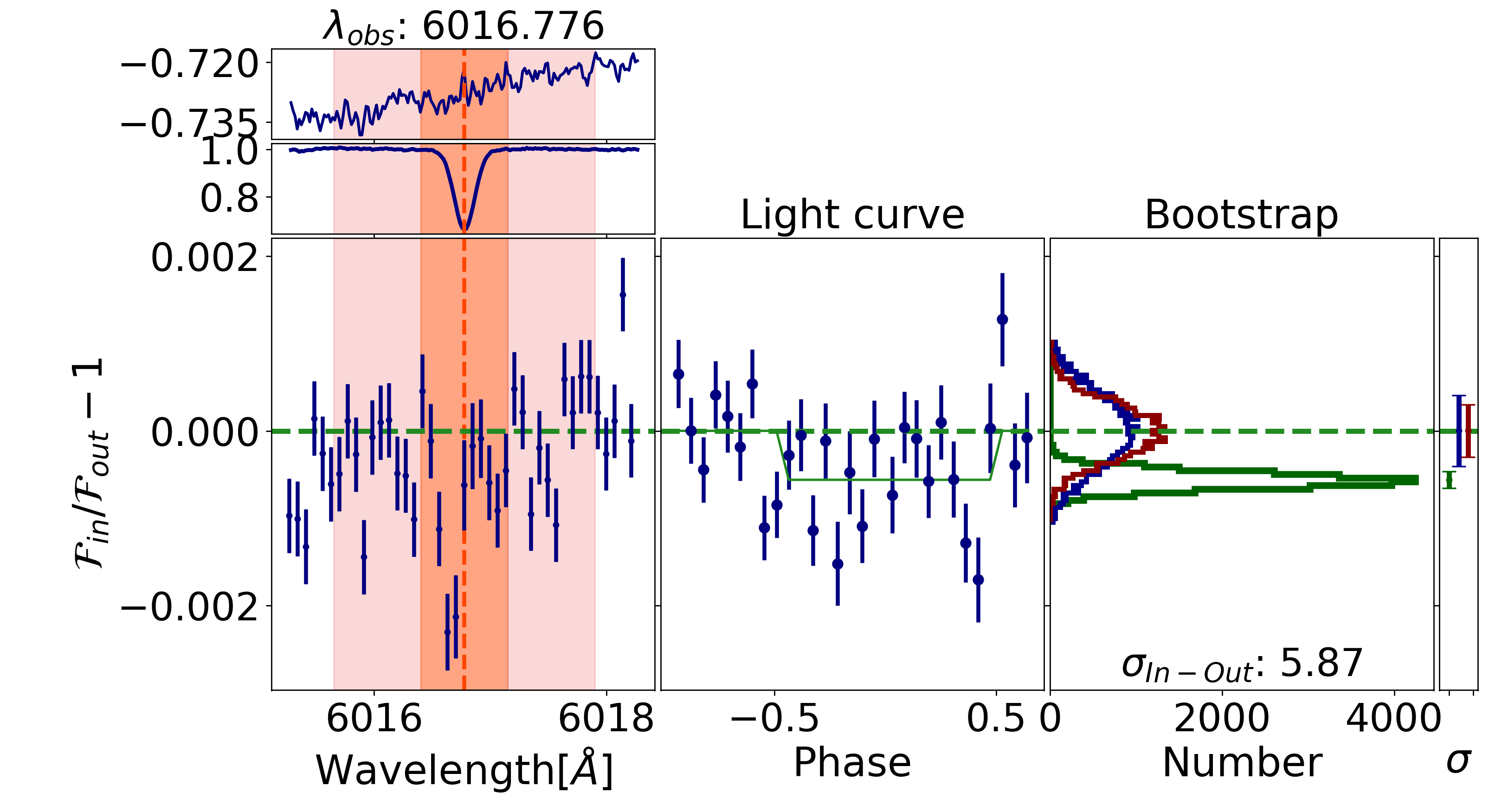}
    }
    \vfill
    \subfloat{
      \includegraphics[width = 0.7\textwidth]{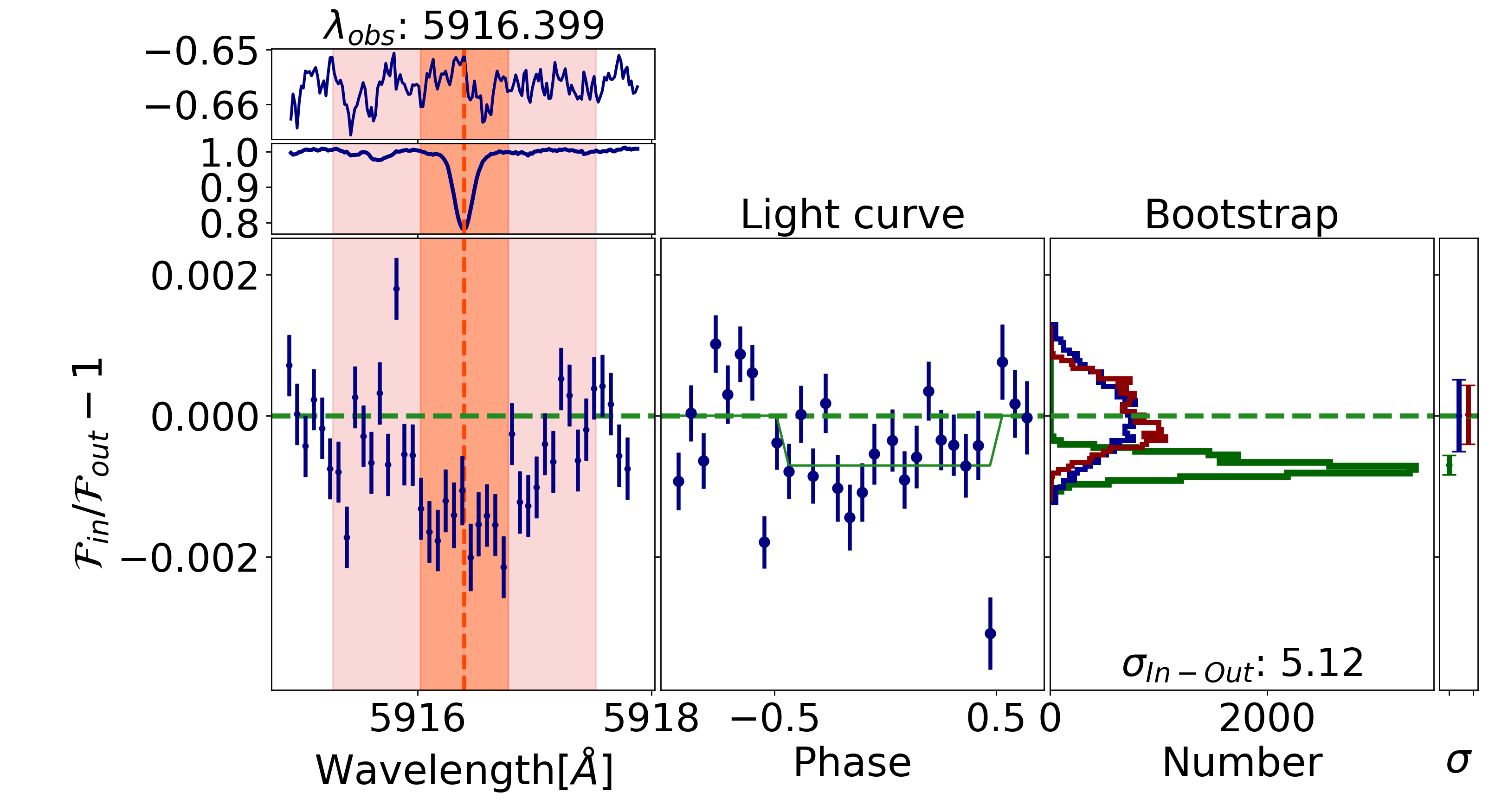}
      }
  
   \caption{Same as Fig.~\ref{fig:valid}, but for Sc II ($\lambda = 6604.6 \angstrom$), Mn I ($\lambda = 6016.7 \angstrom$), and V II ($\lambda = 5916.4 \angstrom $) in HD 209458b}
              \label{fig:thd2094}%
\end{figure*}

The strongest line of Sc II ($6604.6 \angstrom$, see Fig.~\ref{fig:thd2094}) shows absorption with $3.5 \sigma$ of significance, but the one reported in ADR\citeyear{adr2013} ($5526.8 \angstrom$ does not show absorption in our analysis. However, we found the $6604.6 \angstrom$ absorption reliable since its $\Sigma_c$ is very low. Therefore, our result supports the previous detection of Sc~II.

Other works have reported detections of C I, O I, H I, Mg I (\citet{hhd2094}, \citeyear{c&ohd2094}, \citeyear{mghd2094}), and He I \citep{hehd2094alonso}.  \citet{fehd2094} found evidence of Fe II, but did not find Mg I. Additionally, Na I was found by \citet{charbonneau2002}, which has however been recently contested by \citet{rmeffects2020}. \citet{natiohd2094} used ESPRESSO and found that the presence of Na I is still possible, but their data could be explained assuming the sole absorption of TiO, without the existence of Na I. \citet{noatomshd2094} analyzed this planet with the same instrument and found that the absorption, supposedly due to Na I, Mg I, Fe I, Fe II, Ca I, V I, H I, and K I, can be explained by Rositter-McLaughlin effects.

In our analysis, we did not detect any of the species mentioned in the previous paragraph. In the case of C I, the only transition with a detection is shared with Mn I ($6016.7 \angstrom$). The others lines of C I were nondetections (three with a bigger value of the proxy), and the $6016.7 \angstrom$ transition is several orders of magnitude stronger at Mn I in our proxy; therefore, we think that it is likely that the absorption found in our data belongs to Mn I. On the other hand, we did not analyze H I since the transition of this element would need a special tuning of parameters due to its prominent width. The Mg I did not show a detection, but we only have one transition approved by the QT in this spectral range. The only ST of He I did not show absorption. We note that K I has no transitions approved by the QT and O I has no ST.\ Furthermore, Fe I shows several detections and nondetections, but they do not show a correlation with our proxy, nor with the continuum noise; thus, we classified this element as inconclusive, as we could not explain the nondetections.

We highlight the firm nondetection of S I on this dataset because it is consistently nondetected among all transitions analyzed in this spectral range. Although several other elements show nondetections, this is the only one with more than four transitions approved by our QT.

\subsection{HD 189733b (UVES at VLT, HARPS at LaSilla)}

    This planet has some previously reported detections of H I \citep{hhd1897}, He I \citep{hehd1897}, Na I \citep{nahd1897}, and O I \citep{ohd1897}. No spectral lines of H I or O I were accepted by our QT. We note that Na I has only one transition approved by the QT on each instrument, which was not sufficient to conclude that there is absorption, nor to deny it. Furthermore, Na D were rejected by the QT due to their prominent width. Additionally, He I had several more transitions accepted in the UVES data, but it exhibited no consistent absorption in either dataset. Therefore, we could neither confirm nor deny any of these previous detections. In fact, the analysis of the HARPS dataset did not reveal consistent absorption for any element, placing a limit on the S/N requirements of our methodology.
    
\begin{figure*}[t]
    \centering
    \subfloat{
    \includegraphics[width = 0.5\textwidth]{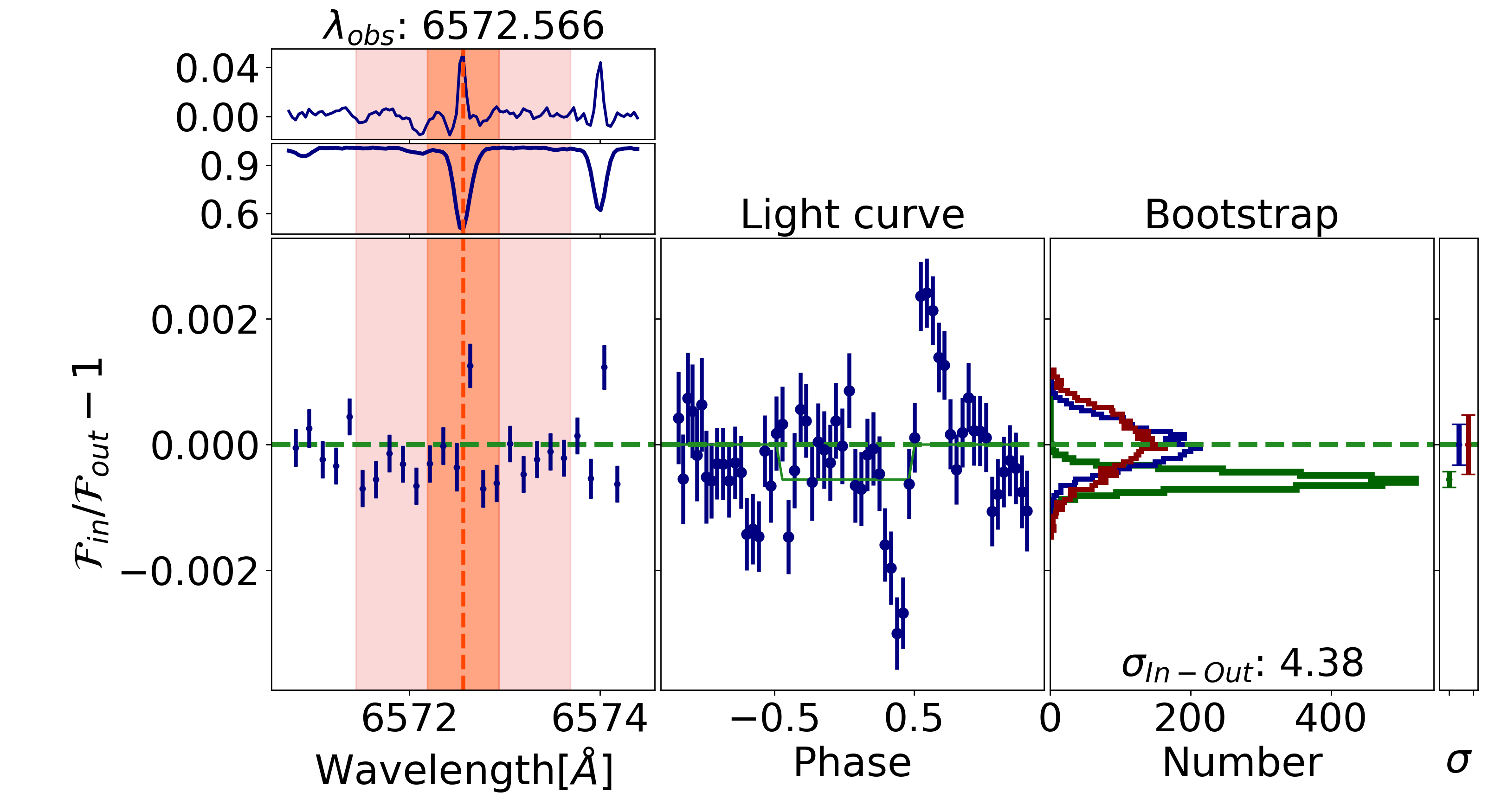}
    
    \includegraphics[width = 0.5\textwidth]{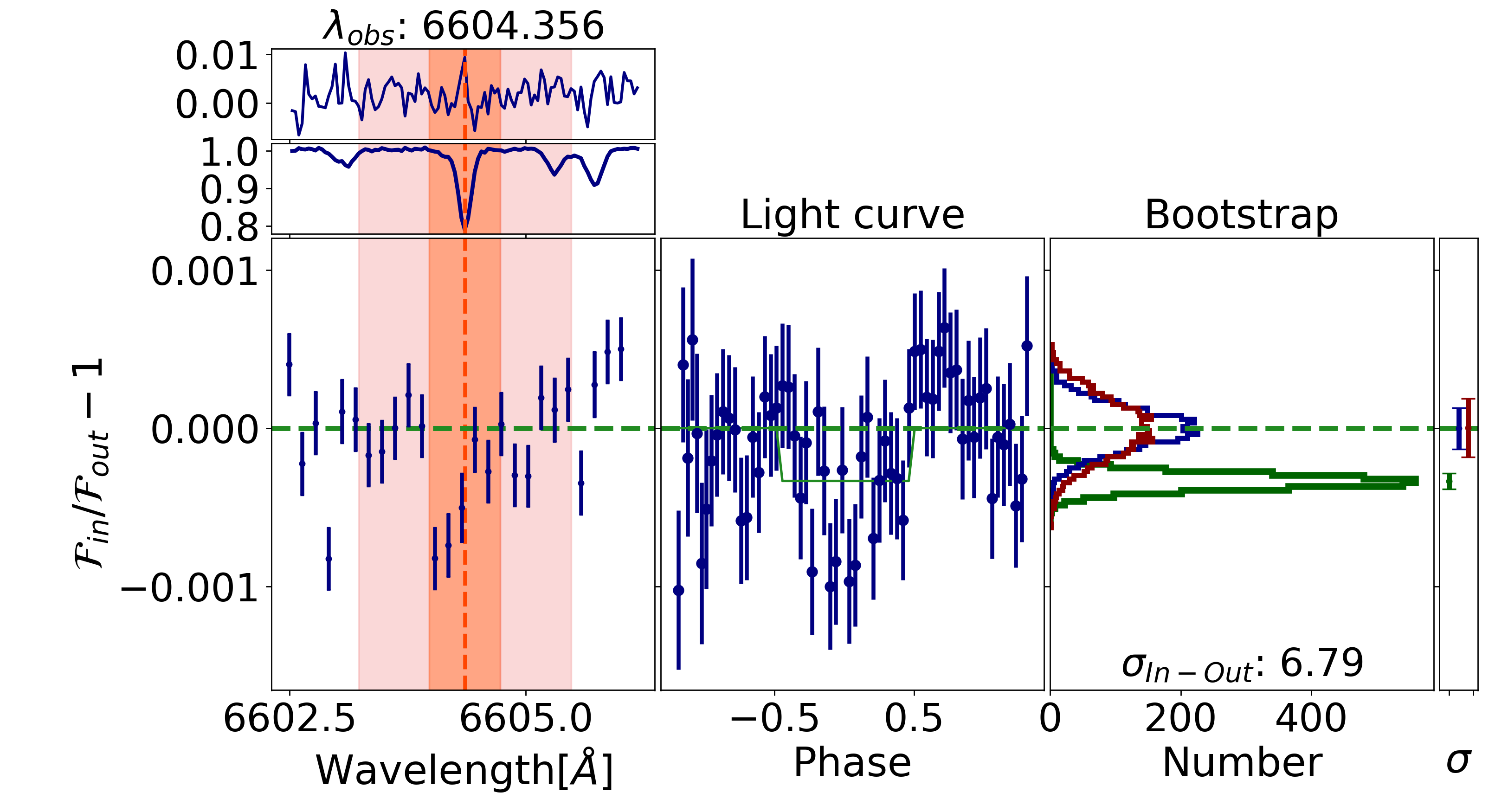}
    }
    \vfill
    \subfloat{
    \includegraphics[width = 0.5\textwidth]{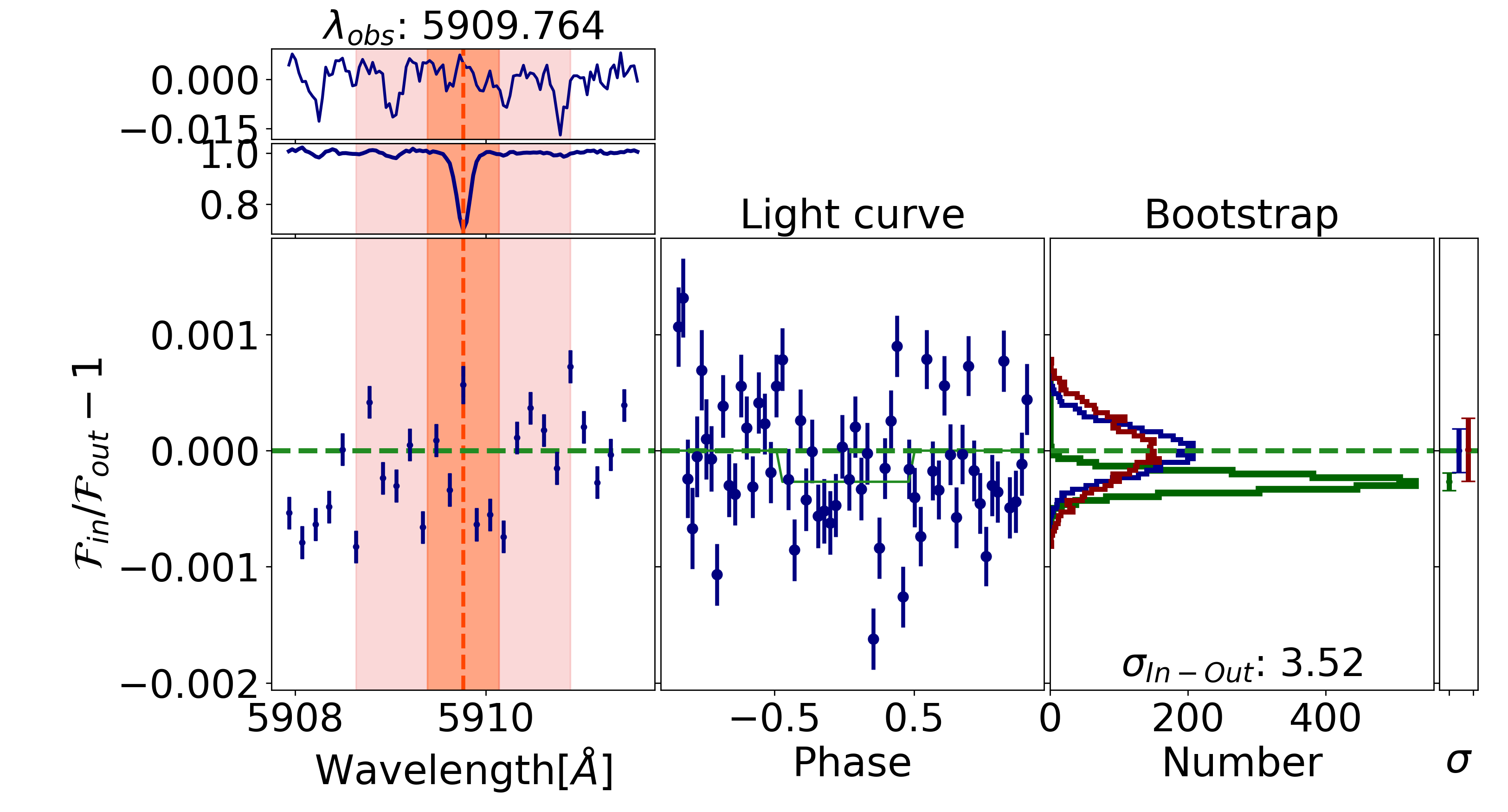}
    
    \includegraphics[width = 0.5\textwidth]{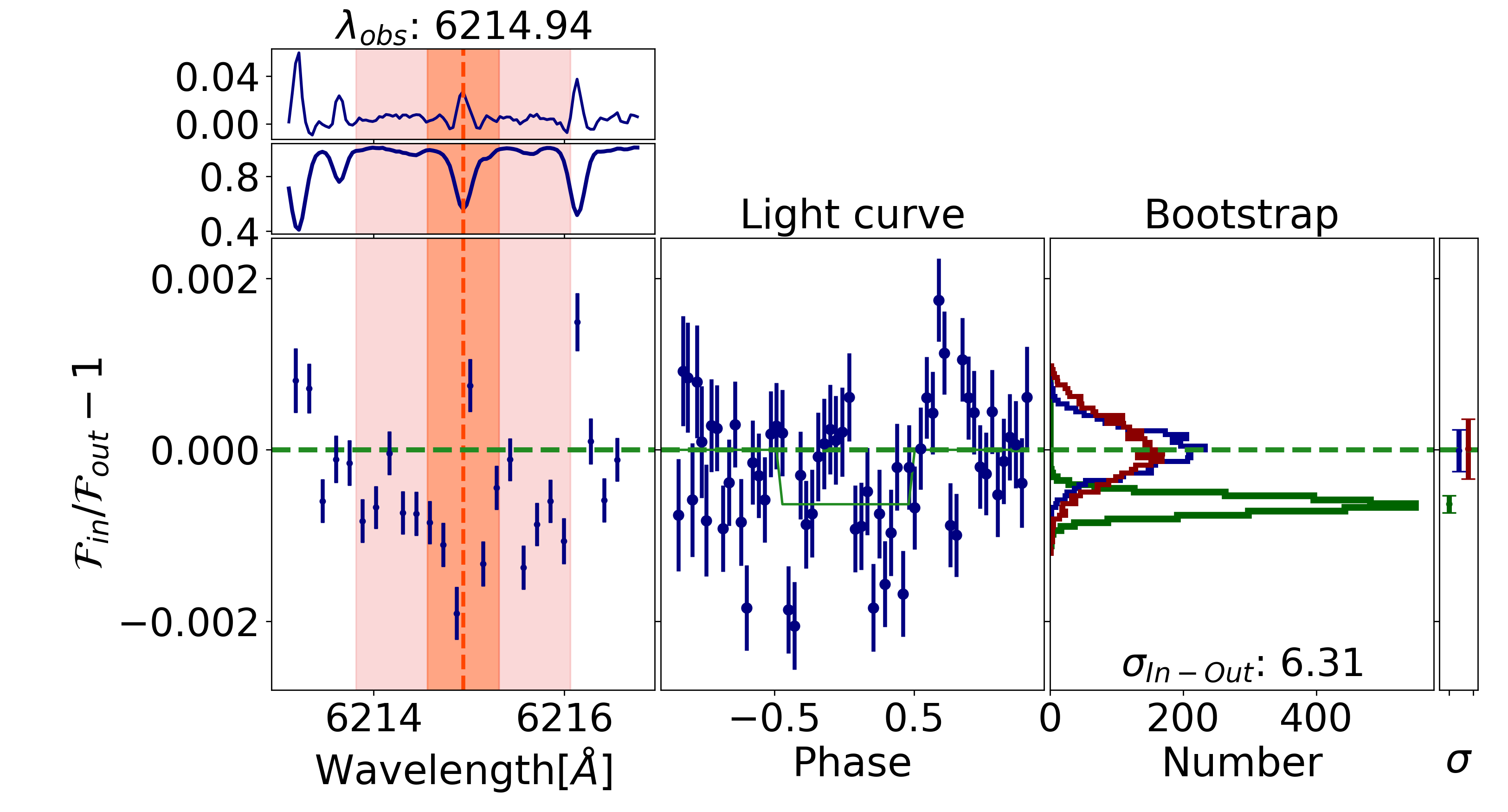}
    }
   
   \caption{Same as Fig.~\ref{fig:valid}, but for Ca I ($\lambda = 6572.8 \angstrom$), Sc II ($\lambda = 6604.6 \angstrom$), and Ti II ($\lambda = 5910.1 \angstrom, ~ 6214.6 \angstrom$) in the UVES dataset of HD 189733b.}
              \label{fig:thd1897}%
\end{figure*}

    In UVES data, we highlight a tentative detection at the $0.75 \angstrom$ passband for the strongest line of Ca I ($4.4\sigma$ at $6572.8 \angstrom$), Sc II ($6.8\sigma$ at $6604.6 \angstrom$), and Ti II ($3.5\sigma$ at $5910.1 \angstrom$ and $6.3\sigma$ at $6214.6 \angstrom$, see Fig.~\ref{fig:thd1897}). However, we must point out that many of the transitions indicative of detection also presented too much structure in their continuum bands or in their telluric spectrum. For that reason, they should all be revisited with better data in future searches on this planet.
    
\subsection{WASP-74b (UVES at VLT) }

Although \citet{obliquitywasp74} searched for atomic absorption on this planet, no elements have been proven so far.\ However,  \citet{tiowasp74} suggested the presence of TiO and VO. 

\begin{figure*}[t]
    \centering
   \subfloat{
   \includegraphics[width = 0.7\textwidth]{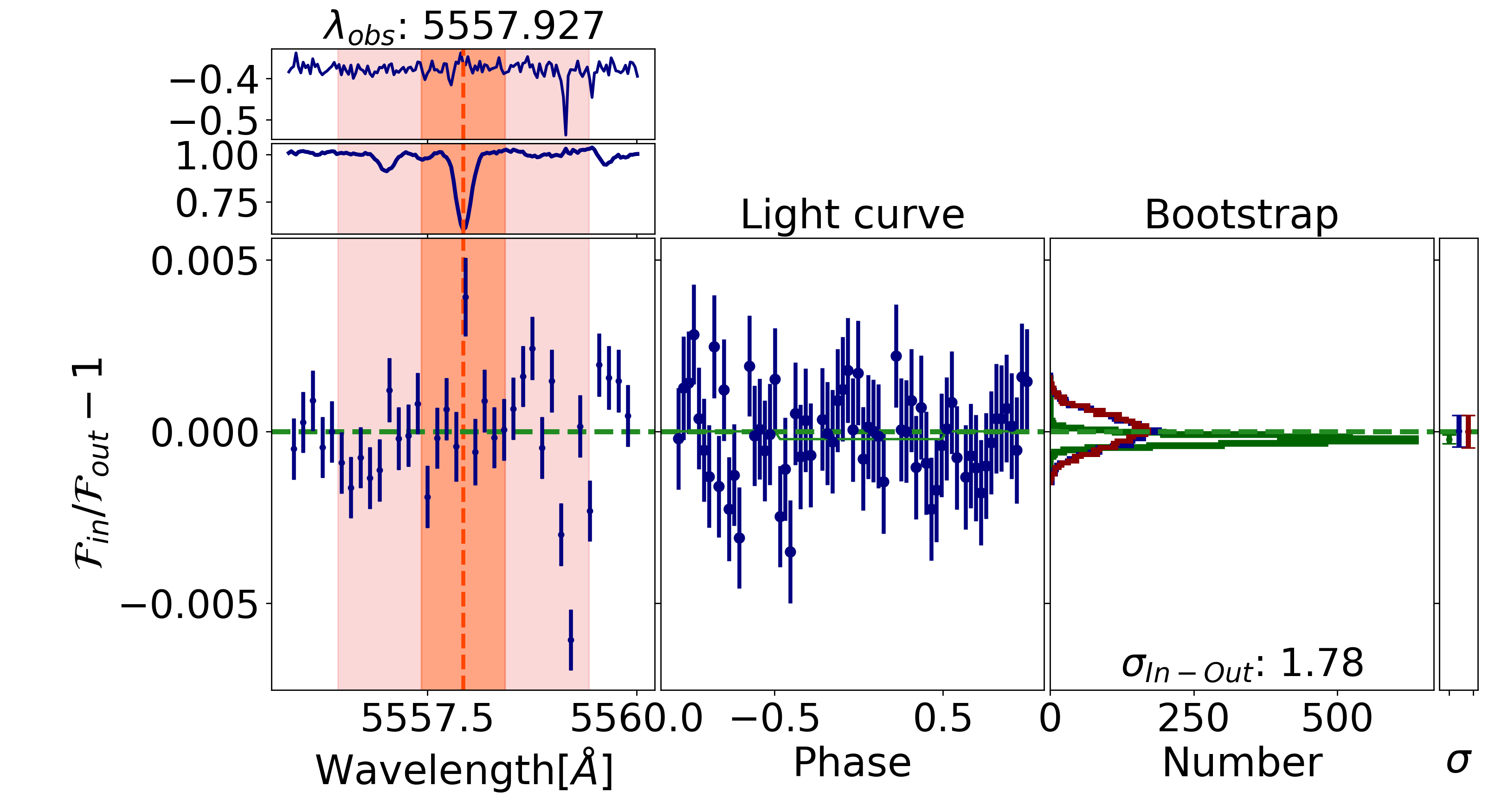}
   }
    \vfill
    \subfloat{
      \includegraphics[width = 0.7\textwidth]{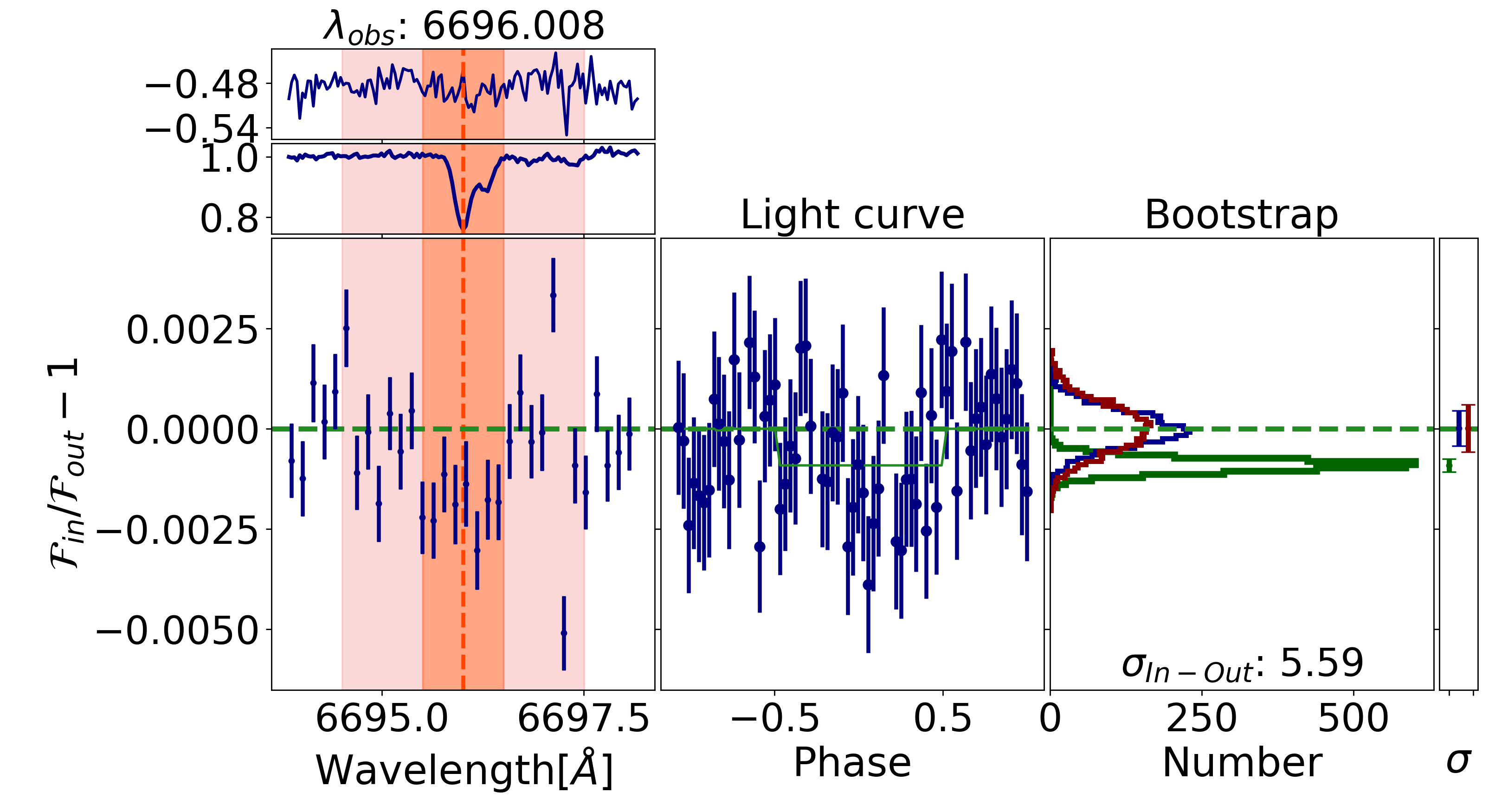}
    }
    \vfill
    \subfloat{
      \includegraphics[width = 0.7\textwidth]{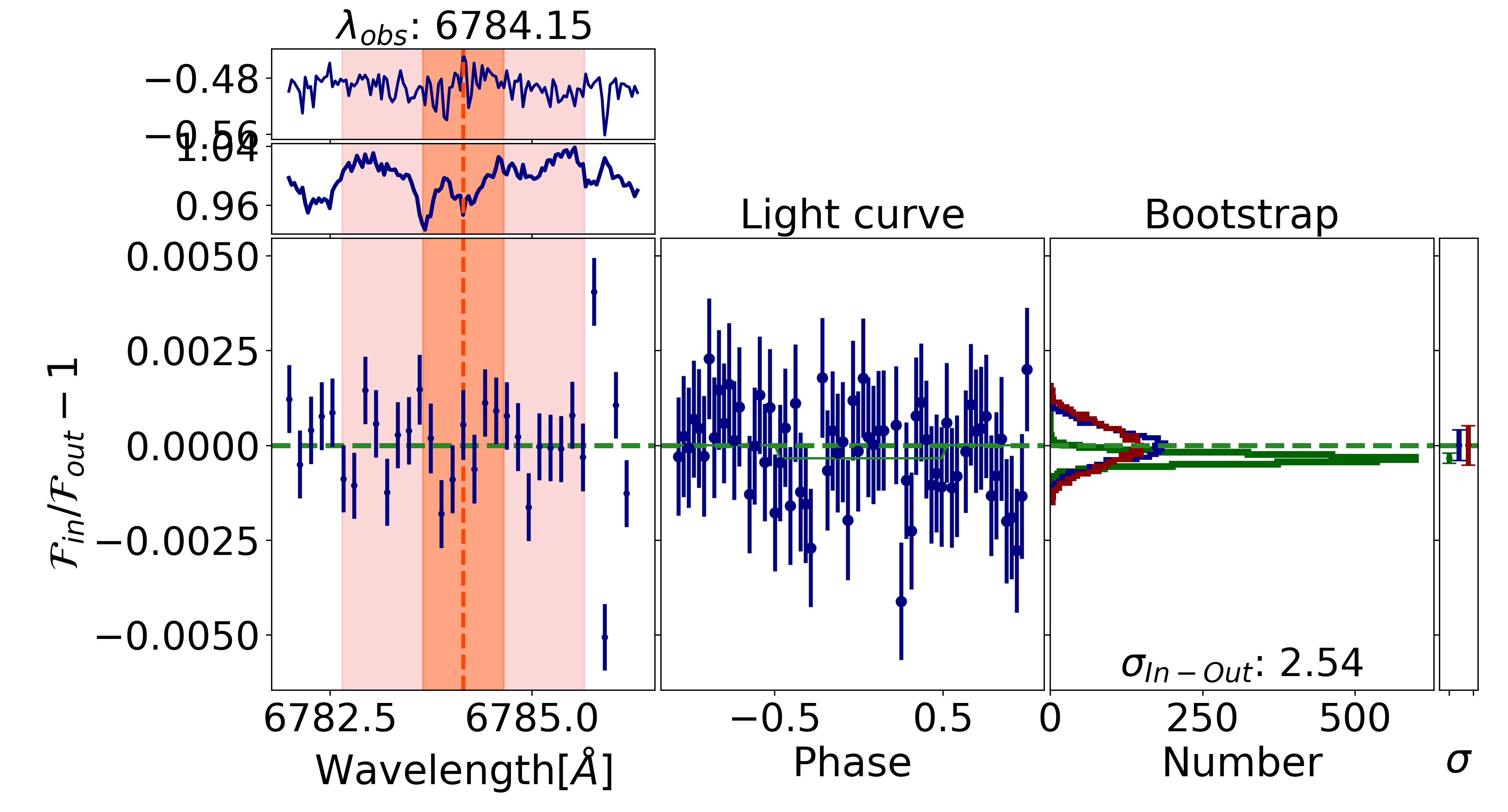}
      }
  
   \caption{Same as Fig.~\ref{fig:valid}, but for Al I ($\lambda = 5557.9, ~ 6696.0, ~6784.3 \angstrom $) in WASP-74b.}
              \label{fig:twasp74}%
\end{figure*}

For this planet we only highlight a tentative detection of Al I after analyzing the four strongest lines. Of these, the strongest and the weakest ($5557.9 \angstrom$ and $6784.3 \angstrom$) did not show absorption; however, they presented structure in its continuum passbands. The other two were very close ($6696.0 \angstrom$ and $6696.2 \angstrom$, see Fig.~\ref{fig:twasp74}). When these two lines were analyzed together with the $1 \angstrom$ passband, absorption was detected with a $5.6 \sigma$ of significance. 

\subsection{TrES-2b and HAT-P-2b (HIRES at Keck)}

\par These planets have a S/N per frame considerably lower than that of HD 209458b. For this reason, we did not find a robust detection of any element in these planets, even with HIRES at Keck.


\section{Conclusions\label{conclusions}}

We present a blind-search methodology designed to perform spectral surveys to search for atomic species in transiting exoplanets. In our work we revisited and expanded on the analysis of the dataset in which \citet{adr2013} discovered Ca I and tentatively Sc II in HD 209458 b. We recovered some of their Ca I detections in some transitions, but found several others of the same element that did not show relative absorption. Therefore, we classify their Ca I detections as inconclusive. On the other hand, our analysis supports the Sc II detection and highlights the possible existence of V II and Mn I in the same planet.

Additionally, we present possible absorption evidence of Ca I, Sc II, and Ti II in the UVES dataset of HD 189733b and Al I in the UVES dataset of WASP-74b. However, we found prominent structures in the retrieved telluric spectrum with this instrument that might be modeled with new telluric correction tools \citep[e.g.,][]{molecfit}. Additionally, we did not find any conclusive detection of any element in the HARPS dataset of HD 189733b, nor in the HIRES datasets of TrES-2b or HAT-P-2b.

Recently, \citet{rmeffects2020} has contested the sodium detection of \citet{charbonneau2002}, measuring the Rositter-McLaughlin and center to limb variation effects on tranists of HD 209458b. \citet{noatomshd2094} used ESPRESSO and found that the apparent signal of several species (including Na I and Ca I) could only be explained with Rositter-McLaughlin effects. These effects could be weak or strong depending on the planet and spectral line analyzed \citep{nawasp52, nawasp166} and they have not been quantified for an algorithm as ours. Therefore, for a more definitive conclusion, it will be necessary to correct for these effects in our analysis.

It must be noted that there exists a uncertainty in the reported values of spectral features. \citet{brass2018} found important differences between databases; for instance, they noticed that the oscillator strength might differ as much as $\pm 4$ dex. This is very concerning to astronomy as these errors could propagate throughout the field, particularly to us since our proxy strongly depends on oscillator strength, and every uncertainty in these values would provoke a big difference in our results. We chose Spectroweb to remain consistent with previous works and complemented it with the Helium lines of VALD.

Even though we could only identify a few possible detections of atomic species with consistent absorption, forthcoming facilities such as the James Webb Space Telescope (JWST) and the Extremely Large Telescope (ELT) will provide data with a considerably higher S/N. For example, our Al I tentative absorption in the UVES dataset of HD 189733b shows a $\sim$6$\sigma$ relative absorption. If we were to observe the same transition with the High Resolution Spectrograph on the ELT (EELT-HIRES), the certainty would theoretically  be approximately five times bigger, under similar conditions, thus providing a powerful technique to model-independently discover and constrain the overall composition of exoplanetary atmospheres.


\footnotesize{ \textit{Acknowledgements:} \label{ack} ALB and RAM acknowledge support from CONICYT/FONDECYT Grant Nr. 1190038. PMR acknowledges support from the Chilean Centro de Excelencia en Astrofisica y Tecnologias Afines (CATA) BASAL PFB/06.\\
This research has made use of data obtained from the ESO Science Archive Facility under request numbers alira524434, alira357282 and alira377917, based on observations made with ESO Telescopes at the La Silla and Paranal Observatories under programme ID's 089.D-0701 A, 072.C-0488(E), and 099.C-0618 respectively.\\
This research has made use of the Keck Observatory Archive (KOA), which is operated by the W. M. Keck Observatory and the NASA Exoplanet Science Institute (NExScI), under contract with the National Aeronautics and Space Administration.\\
Additionally, this work has made use of the VALD database, operated at Uppsala University, the Institute of Astronomy RAS in Moscow, and the University of Vienna. \\
We also thank profusely the referees for their insightful comments.}



\bibliographystyle{aa}
\bibliography{references}

\clearpage

\onecolumn
\appendix

\section{Detailed results of transtions in HD 209458b datasets.}
Here we show the detailed results of the transitions of the elements mentioned in this work. For the tables below, we show, from left to right, 
the element to which the transitions belong, the laboratory wavelength (and the figures in which the observed differs), the value of our proxy and the normed depth (i.e., the difference between the continuum and the minimum of the flux, after normalization). Additionally, the $\sigma_{In-Out}$ column shows the distance from the In-Out center to zero, divided by the width of the In-Out histogram. In contrast, $\sigma_{rel}$ shows the distance between the in-out center to the center of the noise histogram with the maximum thickness, and we took the width of the latter to normalize. To match with plots as in figure \ref{fig:atom_hd20}, the values of absorption are shown as positive.

\begin{longtable}{cccccccccc}%
\caption{Detailed results of transtions in HD 209458b dataset.\label{tab:elhd20}}\\
\hline
\hline
Atom& &\ensuremath{\lambda_{lab} (\lambda_{obs})}&
\ensuremath{ \log{ \left( \frac{gf}{m} \times \exp \frac{-E_{low}}{k_BT} \right) }}
&normed depth&\ensuremath{\sigma_{In-Out}}&\ensuremath{\sigma_{rel}}&\ensuremath{\Sigma_c}&$C_{IO} \times 10^4$&$\Sigma_{IO} \times 10^4$\\%
\hline
\endfirsthead
\caption{Continued.} \\
\hline
Atom& &\ensuremath{\lambda_{lab} (\lambda_{obs})}&
\ensuremath{ \log{ \left( \frac{gf}{m} \times \exp \frac{-E_{low}}{k_BT} \right) }}
&normed depth&\ensuremath{\sigma_{In-Out}}&\ensuremath{\sigma_{rel}}&\ensuremath{\Sigma_c}&$C_{IO} \times 10^4$&$\Sigma_{IO} \times 10^4$\\%
\hline
\endhead
\hline
\endfoot
\hline
\endlastfoot

\multirow{10}{*}{Ca I}&\ensuremath{\bigtriangleup}&6572.779(.942)&{-}4.24&0.08&2.11&0.62&0.004&{-}3.3&1.6\\%
&\ensuremath{\bigcirc}&6162.173(.331)&{-}0.09&0.64&5.52&1.37&0.009&{-}4.7&0.8\\%
&\ensuremath{\bigtriangleup}&6439.075(.235)&0.39&0.58&{-}1.4&{-}0.25&0.005&1.5&1.1\\%
&\ensuremath{\bigtriangleup}&6462.567(.734)&0.26&0.59&{-}1.53&{-}0.45&0.005&2.3&1.5\\%
&\ensuremath{\bigtriangleup}&5598.48(.609)&{-}0.09&0.56&0.95&0.31&0.011&{-}1.2&1.3\\%
&\ensuremath{\bigtriangleup}&5598.48(.605)&{-}0.09&0.56&1.48&0.37&0.008&{-}1.8&1.2\\%
&\ensuremath{\bigcirc}&6493.781(.945)&{-}0.11&0.49&{-}0.0&{-}0.0&0.007&0.0&1.0\\%
&\ensuremath{\bigtriangleup}&5601.277(.404)&{-}0.52&0.46&0.37&0.09&0.01&{-}0.5&1.5\\%
&\ensuremath{\bigcirc}&6471.662(.812)&{-}0.69&0.38&2.8&0.89&0.004&{-}3.4&1.2\\%
&\ensuremath{\bigtriangleup}&5857.451(.599)&0.24&0.52&4.3&0.84&0.006&{-}3.7&0.9\\%
&\ensuremath{\bigtriangleup}&6717.681(.84)&{-}0.52&0.37&2.91&0.95&0.007&{-}5.3&1.8\\%

\hline

\multirow{5}{*}{Mn I}&\ensuremath{\bigtriangleup}&6021.803(.937)&0.03&0.35&2.23&0.56&0.004&{-}2.0&0.9\\%
&\ensuremath{\bigtriangleup}&6016.673(.776)&{-}0.22&0.34&5.87&1.38&0.004&{-}5.6&1.0\\%
&\ensuremath{\bigtriangleup}&6013.513(.63)&{-}0.25&0.29&{-}2.65&{-}0.58&0.004&2.8&1.1\\%
&\ensuremath{\bigtriangleup}&5816.837(.51)&{-}1.04&0.31&1.09&0.35&0.006&{-}1.3&1.2\\%
&\ensuremath{\bigtriangleup}&5816.837(.512)&{-}1.04&0.31&{-}1.66&{-}0.45&0.005&2.2&1.3\\%

\hline

\multirow{5}{*}{Sc II}&\ensuremath{\bigtriangleup}&6604.601(.741)&{-}1.31&0.17&3.45&0.98&0.004&{-}4.7&1.4\\%
&\ensuremath{\bigtriangleup}&5526.79(.936)&0.02&0.42&2.73&0.86&0.011&{-}3.1&1.1\\%
&\ensuremath{\bigcirc}&6245.637(.761)&{-}1.03&0.18&4.46&0.64&0.008&{-}5.8&1.3\\%
&\ensuremath{\bigtriangleup}&5641.001(.118)&{-}1.13&0.22&1.79&0.53&0.011&{-}1.6&0.9\\%
&\ensuremath{\bigcirc}&6279.753(.907)&{-}1.26&0.24&1.84&0.55&0.009&{-}2.0&1.1\\%

\hline

\multirow{1}{*}{V II}&\ensuremath{\bigtriangleup}&5916.354(.399)&{-}2.21&0.22&5.12&1.38&0.004&{-}7.0&1.4\\%
\end{longtable}

\section{Detailed results of transtions in HD 189733b (UVES) dataset.}

\begin{longtable}{cccccccccc}%
\caption{Detailed results of transtions in HD 209458b dataset.  \label{tab:elhd18}}\\
\hline
\hline
Atom& &\ensuremath{\lambda_{lab} (\lambda_{obs})}&
\ensuremath{ \log{ \left( \frac{gf}{m} \times \exp \frac{-E_{low}}{k_BT} \right) }}
&normed depth&\ensuremath{\sigma_{In-Out}}&\ensuremath{\sigma_{rel}}&\ensuremath{\Sigma_c}&$C_{IO} \times 10^4$&$\Sigma_{IO} \times 10^4$\\%
\hline
\endfirsthead
\caption{Continued.} \\
\hline
Atom& &\ensuremath{\lambda_{lab} (\lambda_{obs})}&
\ensuremath{ \log{ \left( \frac{gf}{m} \times \exp \frac{-E_{low}}{k_BT} \right) }}
&normed depth&\ensuremath{\sigma_{In-Out}}&\ensuremath{\sigma_{rel}}&\ensuremath{\Sigma_c}&$C_{IO} \times 10^4$&$\Sigma_{IO} \times 10^4$\\%
\hline
\endhead
\hline
\endfoot
\hline
\endlastfoot
\multirow{10}{*}{Ca I}&\ensuremath{\bigtriangleup}&6572.779(.566)&{-}4.24&0.49&4.38&1.17&0.011&{-}5.5&1.3\\%
&\ensuremath{\bigtriangleup}&6439.075(8.827)&0.39&0.8&2.24&0.61&0.012&{-}6.7&3.0\\%
&\ensuremath{\bigtriangleup}&6462.567(.354)&0.26&0.79&2.27&0.65&0.011&{-}5.2&2.3\\%
&\ensuremath{\bigtriangleup}&6493.781(.564)&{-}0.11&0.72&4.72&1.32&0.017&{-}12.4&2.6\\%
&\ensuremath{\bigtriangleup}&6499.65(.402)&{-}0.65&0.62&0.87&0.24&0.011&{-}1.0&1.1\\%
&\ensuremath{\bigcirc}&6471.662(.444)&{-}0.69&0.64&{-}0.2&{-}0.06&0.009&0.4&2.1\\%
&\ensuremath{\bigtriangleup}&6717.681(.434)&{-}0.52&0.63&1.76&0.48&0.01&{-}2.7&1.5\\%
&\ensuremath{\bigcirc}&6508.85(.633)&{-}2.49&0.17&1.52&0.43&0.008&{-}1.0&0.7\\%
&\ensuremath{\bigcirc}&5867.562(.336)&{-}1.6&0.3&4.83&1.39&0.009&{-}4.3&0.9\\%
&\ensuremath{\bigtriangleup}&5809.118(8.995)&{-}1.07&0.38&{-}5.17&{-}1.46&0.01&5.2&1.0\\%
\hline
\multirow{1}{*}{Sc II}&\ensuremath{\bigtriangleup}&6604.601(.356)&{-}1.31&0.21&6.79&1.82&0.007&{-}3.4&0.5\\%
\hline
\multirow{2}{*}{Ti II}&\ensuremath{\bigtriangleup}&5910.051(9.764)&{-}3.24&0.3&3.52&0.99&0.009&{-}2.7&0.8\\%
&\ensuremath{\bigtriangleup}&6214.6(.94)&{-}2.03&0.44&6.31&1.8&0.013&{-}6.3&1.0\\%

\end{longtable}

\section{Detailed results of transtions in WASP-74b}

\begin{longtable}{cccccccccc}%
\caption{Detailed results of transtions in WASP-74b dataset. Unlike the previous one, here we show the $1 \angstrom$ passband. \label{tab:elwasp74}}\\
\hline
\hline
Atom& &\ensuremath{\lambda_{lab} (\lambda_{obs})}&
\ensuremath{ \log{ \left( \frac{gf}{m} \times \exp \frac{-E_{low}}{k_BT} \right) }}
&normed depth&\ensuremath{\sigma_{In-Out}}&\ensuremath{\sigma_{rel}}&\ensuremath{\Sigma_c}&$C_{IO} \times 10^4$&$\Sigma_{IO} \times 10^4$\\%
\hline
\endfirsthead
\caption{Continued.} \\
\hline
Atom& &\ensuremath{\lambda_{lab} (\lambda_{obs})}&
\ensuremath{ \log{ \left( \frac{gf}{m} \times \exp \frac{-E_{low}}{k_BT} \right) }}
&normed depth&\ensuremath{\sigma_{In-Out}}&\ensuremath{\sigma_{rel}}&\ensuremath{\Sigma_c}&$C_{IO} \times 10^4$&$\Sigma_{IO} \times 10^4$\\%
\hline
\endhead
\hline
\endfoot
\hline
\endlastfoot
\multirow{3}{*}{Al I}&\ensuremath{\bigtriangleup}&5557.948(.927)&{-}2.41&0.4&1.78&0.48&0.012&{-}2.3&1.3\\%
&\ensuremath{\bigtriangleup}&6696.023(.008)&{-}2.85&0.24&5.59&1.55&0.009&{-}9.1&1.6\\%
&\ensuremath{\bigcirc}&6784.256(.15)&{-}1.28&0.04&2.54&0.66&0.012&{-}3.4&1.3\\%
\end{longtable}


\end{document}